\documentclass{article} % For LaTeX2e
\usepackage{MLDD_workshop_2023,times}

\usepackage{amsmath,amsfonts,bm}

\def\eqref#1{equation~\ref{#1}}
\def\1{\bm{1}}

\DeclareMathAlphabet{\mathsfit}{\encodingdefault}{\sfdefault}{m}{sl}
\SetMathAlphabet{\mathsfit}{bold}{\encodingdefault}{\sfdefault}{bx}{n}

\usepackage{hyperref}
\usepackage{url}
\usepackage[pdftex]{graphicx}
\usepackage{multirow}
\usepackage{enumitem}
\usepackage{ragged2e}

\title{Predicting protein stability changes under multiple amino acid substitutions using Equivariant Graph Neural Networks}

\author{Sebastien Boyer, Sam Money-Kyrle \& Oliver Bent \thanks{Corresponding authors: \texttt{\{s.boyer, o.bent\}@instadeep.com}}\\
InstaDeep Ltd, London, UK \\
\texttt{\{s.boyer,s.moneykyrle,o.bent\}@instadeep.com} \\
}

\iclrfinalcopy % Uncomment for camera-ready version, but NOT for submission.
\begin{document}

\maketitle

\begin{abstract}
The accurate prediction of changes in protein stability under multiple amino acid substitutions is essential for realising true in-silico protein re-design. To this purpose, we propose improvements to state-of-the-art Deep learning (DL) protein stability prediction models, enabling first-of-a-kind predictions for variable numbers of amino acid substitutions, on structural representations, by decoupling the atomic and residue scales of protein representations. This was achieved using E(3)-equivariant graph neural networks (EGNNs) for both atomic environment (AE) embedding and residue-level scoring tasks. Our AE embedder was used to featurise a residue-level graph, then trained to score mutant stability ($\Delta\Delta G$). To achieve effective training of this predictive EGNN we have leveraged the unprecedented scale of a new high-throughput protein stability experimental data-set, Mega-scale. Finally, we demonstrate the immediately promising results of this procedure, discuss the current shortcomings, and highlight potential future strategies.
\end{abstract}

\section{Introduction}

Protein stability is a crucial component of protein evolution \citep{godoy-ruiz2004}, it lies at the root of our understanding of many human diseases \citep{peng2016} and plays a major role in protein design and engineering \citep{qing2022}. Protein stability is typically represented as the change in free energy, \(\Delta G\), between the unfolded and folded states \citep{matthews1993} and is a global feature of a protein. A negative \(\Delta G\) of folding indicates an energetically favourable protein conformation; the greater the magnitude of a negative \(\Delta G\), the more stable the conformation. Mutations can alter the favourability of a protein fold, with even single amino acid substitution events potentially disturbing the native conformation of a protein \citep{stefl2013}. For example, a substitution from threonine to methionine in 12/15-Lipoxygenase is a cited potential cause of hereditary cardiovascular diseases \citep{schurmann2011}; the mutation disrupts a chain of stabilising hydrogen bridges, causing structural instability and reducing catalytic activity.
The mutational effect on protein stability is the difference in free energy of folding between the wild type (WT) and mutant proteins, \(\Delta \Delta G\) \citep{matthews1993}. Mutagenic effects on protein stability can be determined experimentally using thermostability assays, with \(\Delta \Delta G\) being inferred from differences between WT and mutant denaturation curves \citep{bommarius2006}. However, these assays are labourious and expensive; to adequately assess mutational effects at a higher throughput rate, researchers have turned to computational methods. The established precedent for computational modelling of mutant stability is empirical physics-informed energy functions, which rely on physical calculations to infer \(\Delta \Delta G\) \citep{marabotti2021}. For example, Rosetta \citep{kellogg2011, das2008} employs Monte Carlo runs to sample multiple protein conformations and predicts folding free energy from physical characteristics. These characteristics of Lennard-Jones interactions, inferred solvation potentials, hydrogen bonding and electrostatics are common to other packages such as FoldX \citep{schymkowitz2005}. While Molecular Dynamics software, such as Amber \citep{case2005}, utilises these characteristics in force fields to explore protein conformational landscapes and calculate potential energies by resolving classical physics calculations.\\
These physics-based models can provide scoring for both protein stability or mutation-induced change of protein stability, however, they are still not fully scalable to large data-sets given the computational expense necessary for each simulated prediction. For example, conformation sampling via Monte Carlo simulations in Rosetta requires extensive compute time. On the other hand, machine learning-based predictors and, more recently, Deep learning (DL) approaches have shown improved scalability and, in some cases, comparable accuracy with physics-based models \citep{iqbal2021}. This work will continue to explore the advantages of an entirely data-driven DL approach for predicting protein stability changes under multiple amino acid substitutions.

\section{Related Work}

In moving away from established molecular modelling approaches, machine learning methods EASE-MM \citep{folkman2016} and SAAFEC-SEQ \citep{li2021} leverage 1D sequences and protein evolutionary information to predict \(\Delta \Delta G\) with decision trees and Support Vector Machines, respectively. While ACDC-NN-Seq \citep{pancotti2021deep} explored utilising DL by applying Convolutional neural networks (CNNs) to protein sequences. As sequence data is more widely available than experimental structures, it is probable that the insight of these models into 3D structural characteristics, such as free energy of folding, is limited by their 1D representation. PON-tstab \citep{yang2018} implemented a combination of sequence and structure-based features in tabular format with random forests. DeepDDG \citep{cao2019} relies on tabular empirical features obtained from structure, such as solvent-accessible surface area, to predict stability with neural networks. However, tabular features engineered from structure are a restrictive depiction of protein geometry; graph-based approaches provide a promising alternative representation, with encouraging results when applied to protein structure prediction \citep{delaunay-etal-2022-gnn}.\\

In particular, three DL models; ThermoNet \citep{li2020predicting}; RASP \cite{blaabjerg2022rapid}; and ProS-GNN \citep{wang2021pros}, have engaged in combining the two physico-scales involved in understanding protein geometry: the atomic scale and the residue scale of interactions. Both ThermoNet and RASP learn a representation of the atomic environment (AE) around the pertinent (mutated) residue using 3D CNNs before passing this representation through a Multi-layer perceptron (MLP) to score the mutational effect on protein stability. While obvious similarities exist between those two models, they are very different at their core.
ThermoNet determines the AE representations on the fly, utilising both WT and simulated mutant structures as inputs for the MLP in the same loop. RASP initially trains a self-supervised AE embedder on a masked amino acid task, then uses this embedding as input features for a coupled WT and mutant amino acid encoding to feed an MLP trained on stability scoring.
Moreover, ThermoNet is trained on a rather small experimental data-set (n $\sim 3,500$), while RASP is trained on a large data-set (n $\sim 10^{6}$ for the AE embedder and n $\sim 2.5 \times 10^{5}$ for scoring) of Rosetta simulated scores, making it an emulator of the physics-based scorer.
The third DL approach, ProS-GNN \citep{wang2021pros}, replaced the CNN ThermoNet atomic environment embedding layer with GNNs. ProS-GNN also shares with ThermoNet and other DL models like ACDC-NN the constraint of being antisymmetric to reversed mutation.
The aforementioned state-of-the-art stability prediction models in the literature share the following caveats:
\begin{enumerate}[topsep=0pt,itemsep=1ex,partopsep=1ex,parsep=1ex]
\item Their underlying architecture allows only single amino acid substitutions.
\item Big experimental data-sets with the necessary structural data for these models are lacking.
\end{enumerate}
Indeed, RASP is constrained to predicting on a fixed number of amino acid substitutions by the MLP scorer, which requires a fixed input shape; additional mutations increase the dimensions of the AE embedding to an incompatible size. In ThermoNet and ProS-GNN, the impossibility of decoupling the atomic and residue scales prohibits multiple amino acid substitutions; the required size of voxel or graph for multiple, even proximal, substitutions would be rapidly unmanageable.\\
A solution for both caveats exists. The self-supervised AE embedder of RASP already decouples the atomic and residue scales, and GNNs allow for some flexibility in graph topology, enabling consideration of multiple residues rather than only the embedding of the residue of interest. Integrating the RASP AE embedder with a graph-based approach would enable scoring of multiple substitution events. On the experimental data front, a new data-set, Mega-scale \citep{Tsuboyama2022}, based on high-throughput protein stability measurements, was published in late 2022. With over 600,000 data points of single and double mutants spanning over 300 WT structures, it provides a consistent (in terms of experimental set-up) and large data-set, with the express purpose of training models to score the effects of single or double mutations on protein stability.
In light of these observations, we contribute a JAX-implemented solution for resolving these constraints using two E(3) equivariant graph neural networks (EGNNs) \citep{satoras2021}. The first EGNN is trained in a self-supervised way. The second is trained on the Mega-scale data set for scoring mutational effects on protein stability.
\section{Method}
\subsection{Atomic Environment (AE) Embedder}
\begin{figure}
\centering
\includegraphics[width=0.99\linewidth]{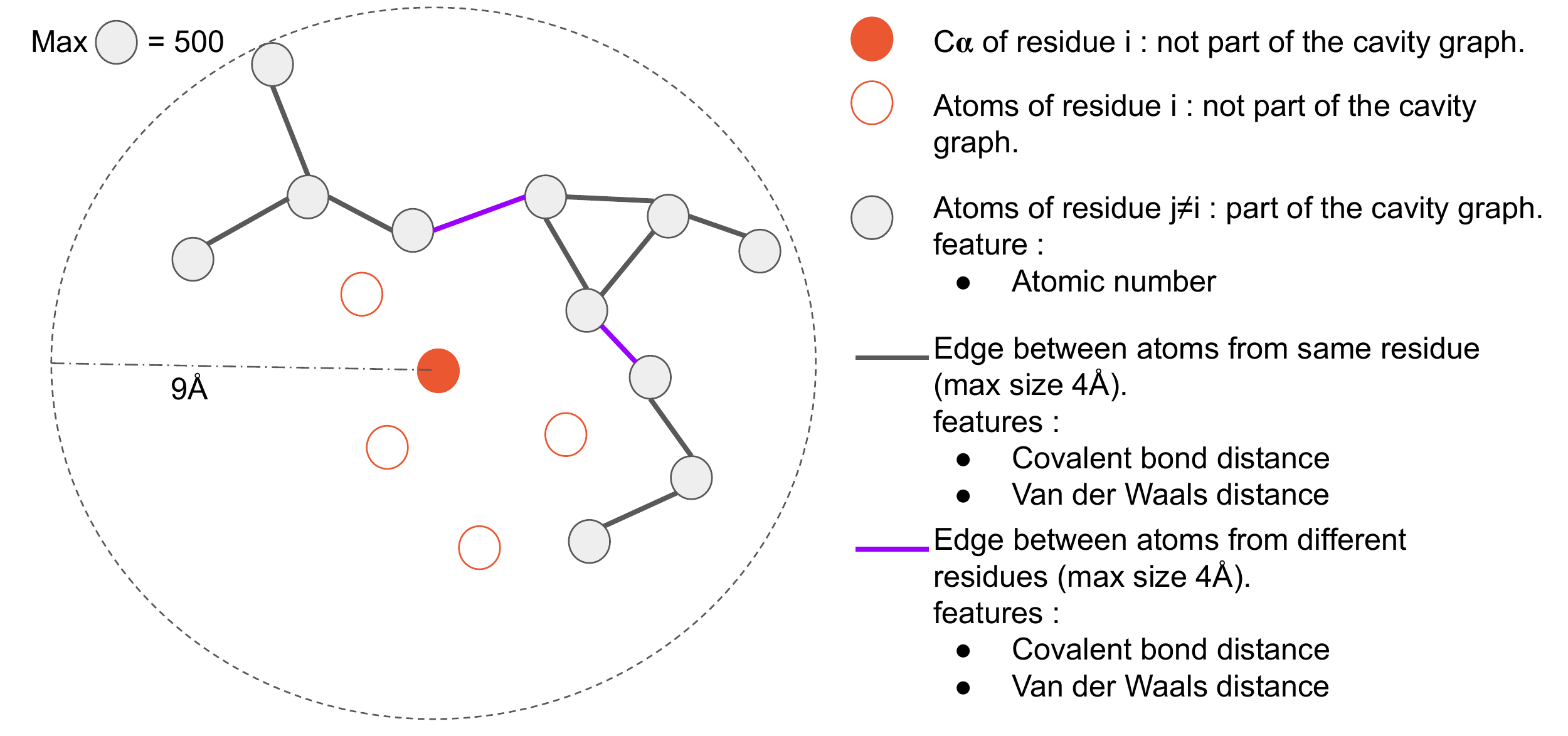}
\caption{Definition of the atomic environment (AE) graph.}
\label{Atomic_env}
\centering
\end{figure}

\begin{figure}
\centering
\includegraphics[width=0.99\linewidth]{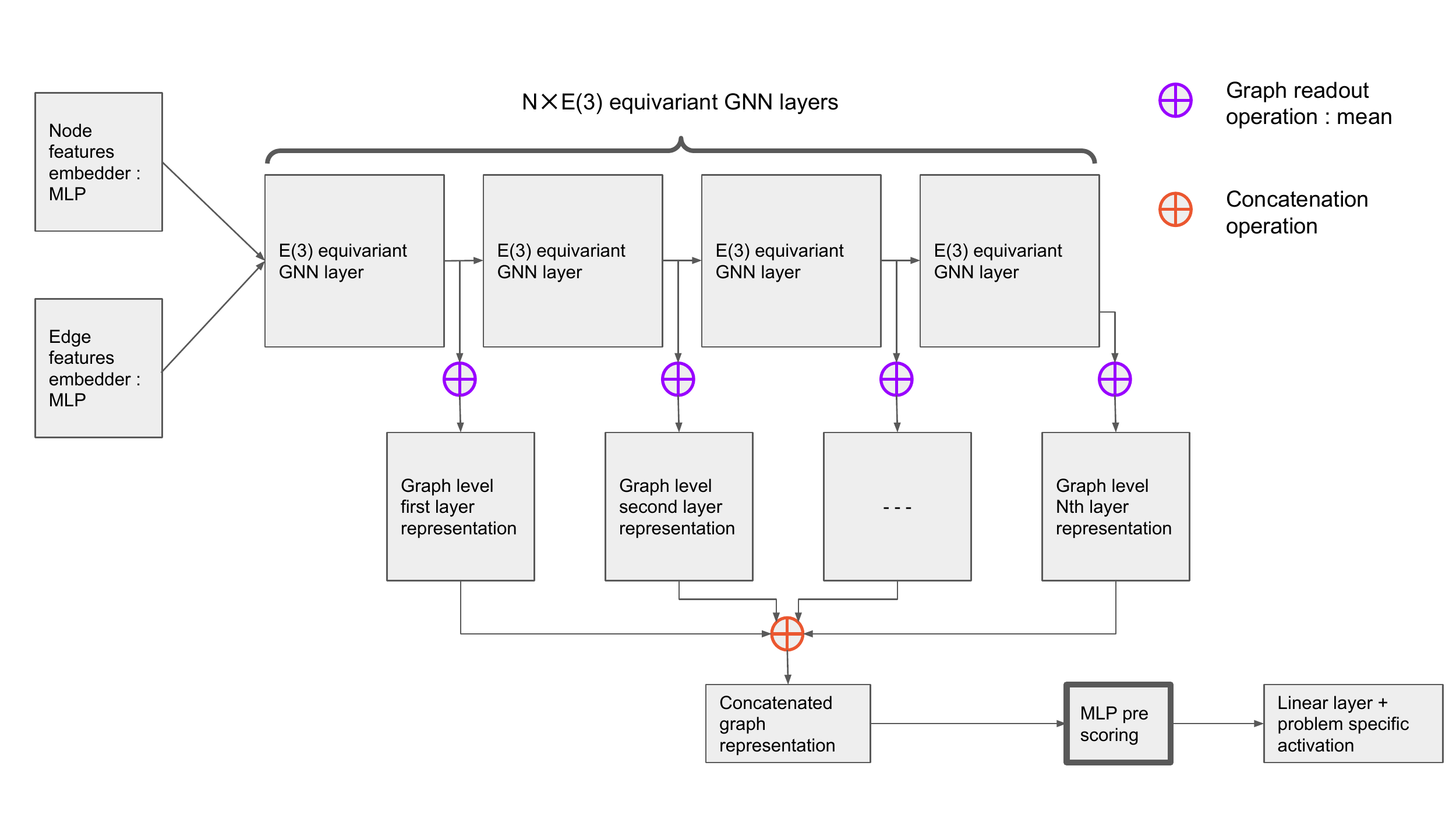}
\caption{Backbone of the E(3) equivariant graph neural network (EGNN) used for both AE embedding and scoring tasks. The EGNN layer is the EGCL taken from \citep{satoras2021}.}
\label{EGNN}
\centering
\end{figure}

We followed the RASP protocol to design and train our AE embedder in a self-supervised masked amino acid manner, with two key differences:
\begin{enumerate}[topsep=0pt,itemsep=1ex,partopsep=1ex,parsep=1ex]
\item We used an EGNN (Figure \ref{EGNN}) with its own set of graph features describing the AE (Figure \ref{Atomic_env}) instead of a CNN.
\item We used a macro averaged F1 score as our metric on the validation set to select model parameters from the highest-performing epoch.
\end{enumerate}
The training and the validation sets are from the same data-set described in RASP \citep{blaabjerg2022rapid}.
Our EGNN was built with layers described in \citet{satoras2021}, with an average message aggregation strategy (Equation \ref{eq: E(n)_model}). Recalling from \citet{satoras2021} that $\bm{h^l}$ are node embeddings at layer $l$ and $\bm{x^{l}}$ are coordinate embeddings at layer $l$ (atoms coordinates), we defined the equivariant graph convolutional layer (EGCL), as they do, up to the use of this $\frac{1}{N^{neighbors}_i}$ coefficient which allows the re-scaling of different messages according to the node of interest's number of neighbors (hence the average). As with their implementation, $\phi_e,\phi_x,\phi_h$ are MLPs, $a_{i;j}$ defines edge features between node $i$ and $j$, and finally, $\mathfrak{N}(i)$ is the set of neighbors of node $i$.
\begin{equation}
\begin{aligned}
\bm{m}_{i,j}&=\phi_e(\bm{h_i^l},\bm{h_j^l},\left\lVert \bm{x_i^l}-\bm{x_j^l}\right\rVert^2,a_{i,j})\\
\bm{x_i^{l+1}}&=\bm{x_i^l}+\frac{1}{N^{neighbors}_i}\times\sum_{j\neq i,j \in \mathfrak{N}(i)}(\bm{x_i^l}-\bm{x_j^l})\phi_x(\bm{m}_{i,j})\\
\bm{m}_{i}&=\frac{1}{N^{neighbors}_i}\times\sum_{j\neq i,j \in \mathfrak{N}(i)}\bm{m}_{i,j}\\
\bm{h_i^{l+1}}&=\phi_h(\bm{h_i^l},\bm{m}_{i})
\end{aligned}
\label{eq: E(n)_model}
\end{equation}
Node embeddings are passed sequentially through each $N$ layer of the network. After each layer, node embeddings are copied, aggregated with an average graph level readout (global mean pooling) Equation \ref{eq: Graph_level_readout}, and saved. Finally, all the graph representations derived from the different layers are concatenated, Equation \ref{eq: Graph_level_readout}, to form the graph-level embedding $\bm{h_G}$ for the AE sub-graph of a residue $G$, and processed through an MLP toward the desired prediction shape.\\
\begin{equation}
\begin{aligned}
\bm{h_G}&=\mathtt{Concat}(\mathtt{Average}(\{\bm{h_i^{l}}\vert i \in G\})\vert l=0,...,N)
\end{aligned}
\label{eq: Graph_level_readout}
\end{equation}
For building the AE graph, we followed part of the RASP protocol:
\begin{itemize}[topsep=0pt,itemsep=1ex,partopsep=1ex,parsep=1ex]
\item We considered only atoms within a 9{\AA} radius of the C$_{\alpha}$ of interest.
\item We removed the atoms that were part of the residue of interest.
\end{itemize}
Nodes are atoms featurised with a single number (atomic number). Edges are drawn between nodes if two nodes are within a 4{\AA} distance. Edges are featurised by a binary label distinguishing whether the edge is intra- or inter-residue, as well as 2 numbers encoding a notion of the typical distance between the two atoms linked by these edges:
\begin{itemize}[topsep=0pt,itemsep=1ex,partopsep=1ex,parsep=1ex]
\item The sum over the two atoms involved in the edge, of their covalent radius.
\item The sum over the two atoms involved in the edge, of their Van der Waals radius.
\end{itemize}

In this particular instance of the model, distances between atoms are not directly encoded as an edge feature, but given the use of an EGCL, this distance is present as a distance vector (Equation \ref{eq: E(n)_model}) (rather than the usual scalar distance, hence the necessity of E(3) equivariance) by design. Finally, we trained the model on a classification task consisting of retrieving the amino acid around which the AE has been built. Model parameters were selected from the epoch with the best macro F1 scores on the validation set.
A detailed description of the model in terms of its hyper-parameters is provided in the Appendix (Table \ref{Table_embedder_model}).\\

\subsection{Mutant Stability Scoring}
We used the same model architecture as presented for the AE embedding (Figure \ref{EGNN}), for the regression task of predicting $\Delta\Delta G$. The set of hyper-parameters differs as described in the Appendix (Table \ref{Table_scorer_model}). For this task, the graph is built at the residue-level with additional atomic-level features to bridge the gap between the two fundamental scales. Indeed, in this representation nodes are residues represented in terms of their spatial positioning by the residue mean atomic position coordinates. Nodes are featurised with the vector output of the previously trained AE embedder. More node features are included with an 11-dimensional representation of the physico-chemical properties of the WT amino acid \citep{kawashima2007, xu2020}, concatenated to the same representation, except for the mutated amino acid nodes. When a particular node is not mutated this concatenation is just twice the WT physico-chemical 11-dimensional representation.\\
At the edge level, an edge is drawn between two nodes if the mean atomic position distance between the two nodes is within 9{\AA}. The graph is centered around the mutant residue and residues are added given the distance threshold up to n (here 1) edges away from the mutant nodes. In the case of multiple mutants, we allow the different graphs centered around their mutant node, to be disconnected from each other. Features for the edges follows a similar strategy to the atomic graph: 
\begin{itemize}[topsep=0pt,itemsep=1ex,partopsep=1ex,parsep=1ex]
\item A single number to stipulate if the two residues are linked by a backbone, bound or not.
\item Two numbers to provide a specific scale for the distance between two WT residues: the sum of the residue side chain sizes, defined as (i) the maximum distance between the C$_{\alpha}$ and any atoms of the residue; (ii) the maximum distance between two atoms from the same residue.
\item The same two numbers are produced for mutants involved in the edges. When there are no mutants involved then they are just duplicated from the WT numbers.
\item The C$_{\alpha}$/C$_{\alpha}$ distance and the mean atomic position/mean atomic position distance.
\end{itemize}
Finally, to help the training while homogenizing the ranges and variance of the target variable (here the experimental $\Delta\Delta G$) we used a Fermi transform, also as described in RASP \citep{blaabjerg2022rapid}. Our loss function is a simple Root Mean Squared Error, and the best model as well as the best epoch is chosen based on the spearman rank correlation coefficient on the validation set.

\begin{figure}[h]
\centering
\includegraphics[width=0.99\linewidth]{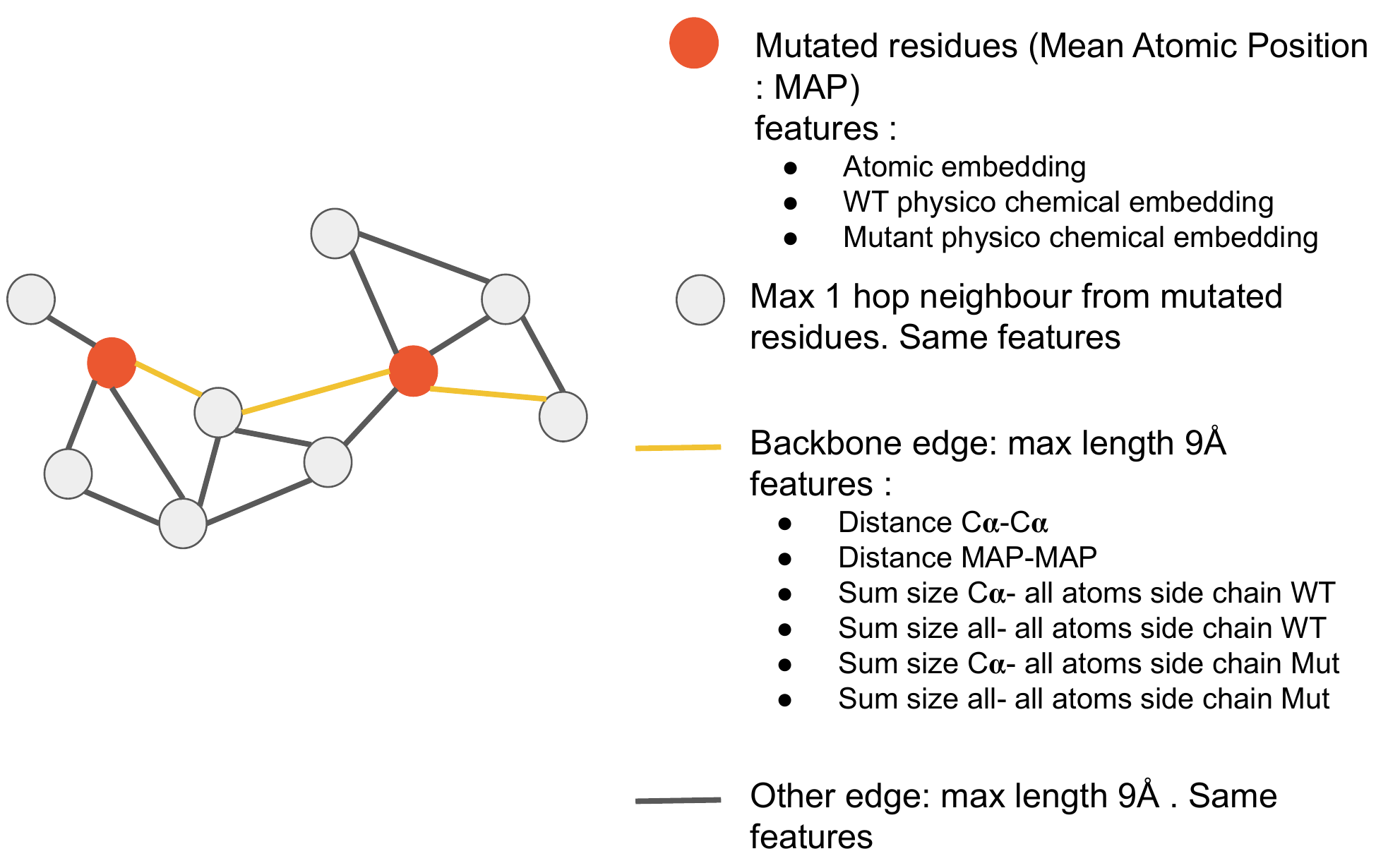}
\caption{Definition of the residue sub-graph build around mutated residues.}
\label{Subgraph}
\centering
\end{figure}

\begin{figure}[h]
\centering
\includegraphics[width=0.99\linewidth]{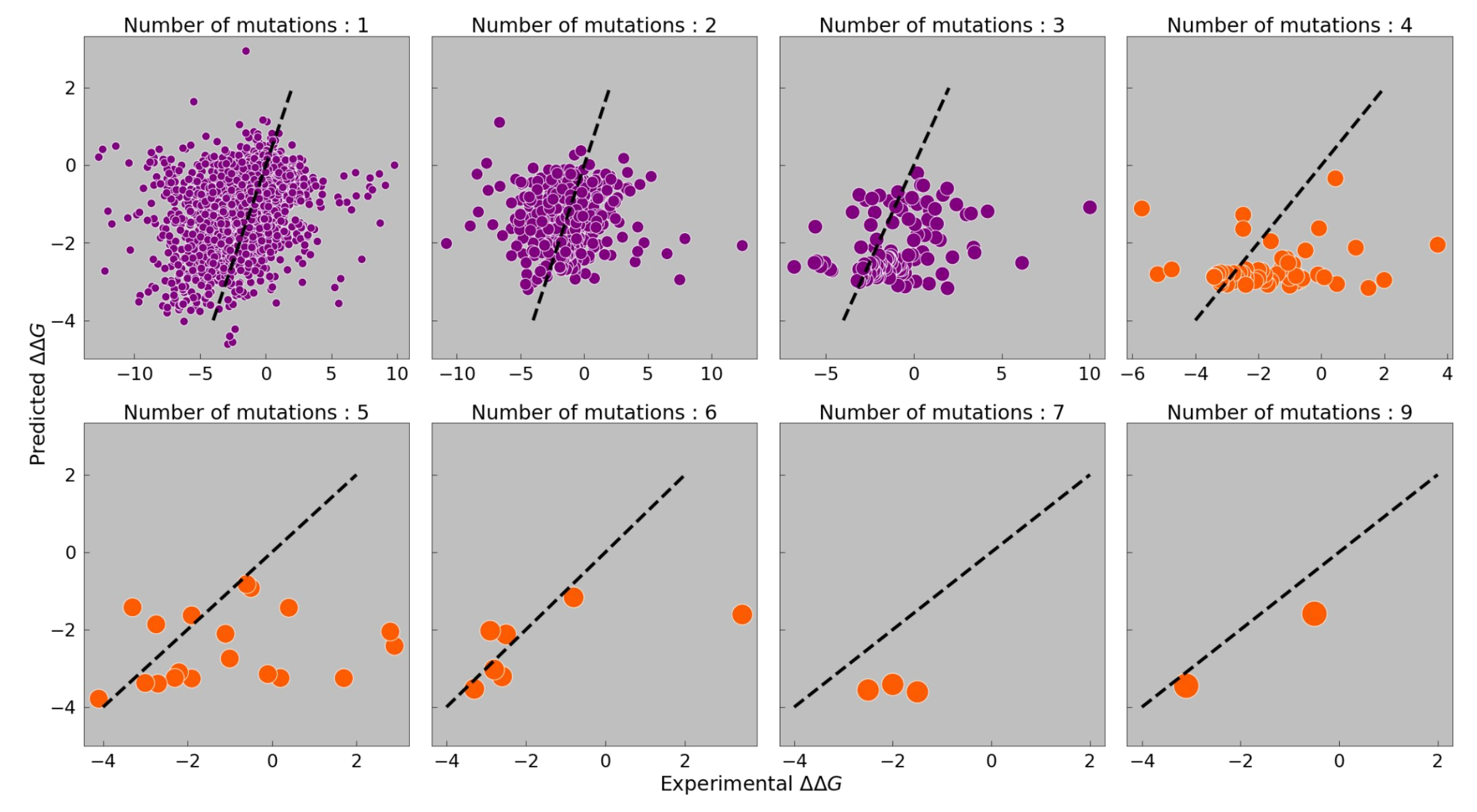}
\caption{Evaluation of our scorer for differing numbers of mutations. Purple markers for Spearman rank correlation p-value$<$0.05 else orange. Marker size is proportional to the number of mutations.}
\label{Num_mut}
\centering
\end{figure}

\section{Results}
\subsection{Atomic Environment Embedder}
With our AE implementation, we reached a macro averaged F1 score of 0.63 on the training set and of 0.61 (accuracy = 0.65) on the validation set, which is comparable to the RASP 0.63 accuracy also on a similar but different validation set (we shuffled the structures), full results in Appendix Table \ref{Evaluation_Embedder}. The confusion matrix on the validation set (Figure \ref{CM_val}) is also provided in the Appendix and shows a variable but strong ability of the model to match ground truth.

\subsection{Mutant Stability Scorer}
\begin{figure}[h]
\centering
\includegraphics[width=0.83\linewidth]{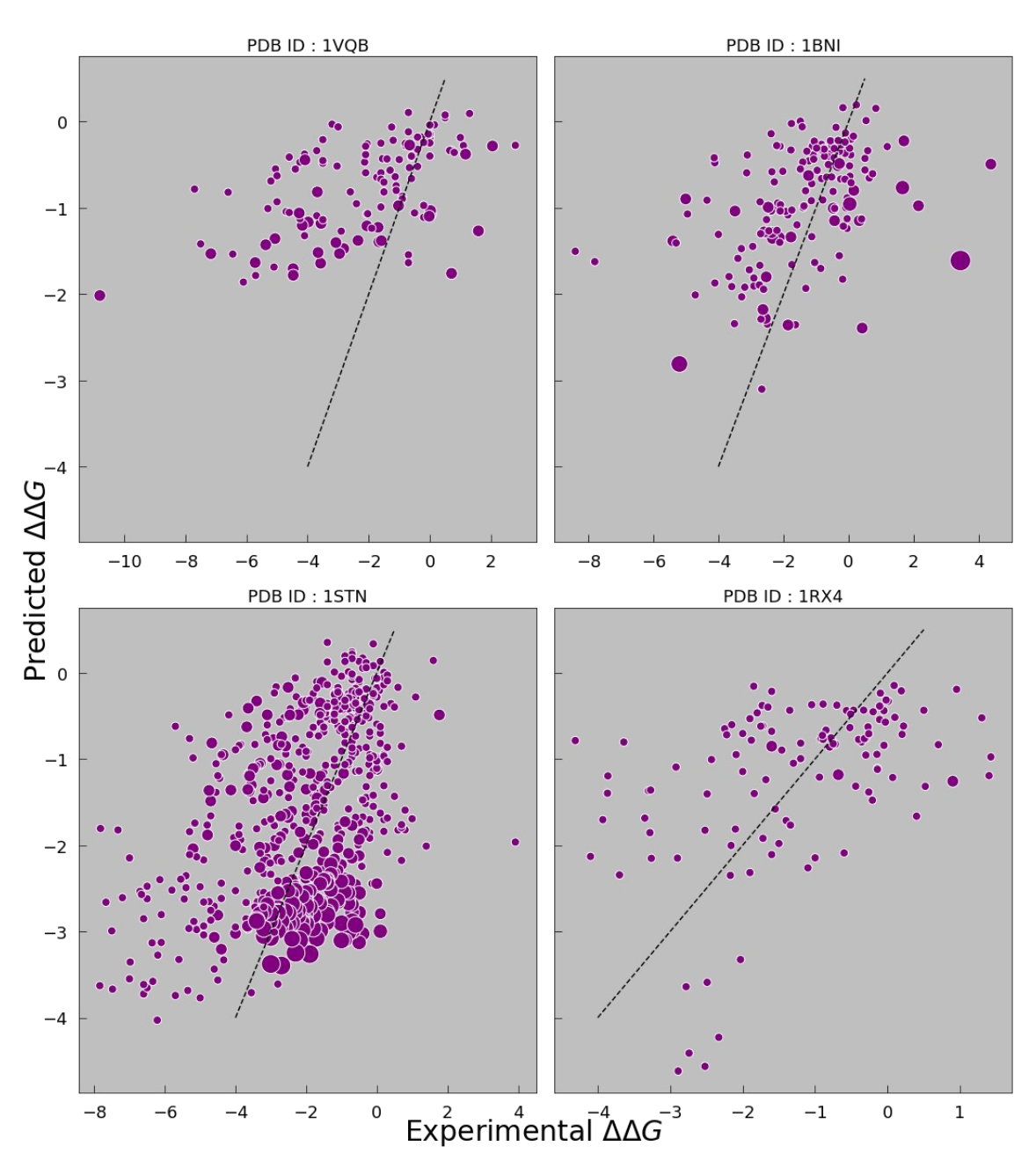}
\caption{Evaluation of our scorer on individual structures, PDBs \cite{burley2017protein}, chosen with at least 100 occurrences in the ThermoMutDB test data-set.  All four structures have a significant prediction correlation (p-value$<$0.05) and the marker size is proportional to the number of mutations in the experiment. Further results breakdown in Table \ref{pdb_mutation_metrics}.}
\label{pdbid}
\centering
\end{figure}

Evaluation metrics for the different splits are available in Figure \ref{Train_val_regress_pred}, as well as a description of the Mega-scale data-set in the Appendix: \ref{sec:appendix}. Given the unique qualities of the Mega-scale data-set, we decided to evaluate the model in what we believe is a more stringent way than simply looking at the Mega-Scale test split (metrics are provided for the split too). Indeed, the Mega-scale data-set only contains domains and not full proteins, and structures were resolved computationally using AlphaFold \citep{jumper2021}. The Mega-scale data-set also only contains up to double mutations. Hence, we decided to evaluate our model on a more standard data-set with experimentally resolved entire protein structures: ThermoMutDB \citep{xavier2021thermomutdb} (a description of the ThermoMutDB data-set is also provided in the Appendix: \ref{app:thermomut}).\\
Over the pooled ThermoMutDB data-set our scorer achieved RMSE = 2.288; Spearmanr = 0.311; Pearsonr = 0.251 Table \ref{Evaluation_scorer}. Interestingly the model seems to generalise well to structures with more than two mutations (Figure \ref{Num_mut}), for which it has not been directly trained. Spearman correlation, for example, spans a range between 0.159 and 0.381, for a number of mutations going from 1 to 3. \\
At the level of individual structures (Figure \ref{pdbid}), model performances can also vary quite drastically from WTs with at least 100 mutants(for an overview per pdbs see Figure \ref{Distri_corr_pdbid}).\\
Finally, we also compared, for a subset of one-point mutations, our work to RASP (Figure \ref{RASP_comp}). On pooled single mutations our proposed approach performs significantly worse (Pearsonr RASP = 0.53; Pearsonr for this work = 0.42 ). Yet our approach outperforms RASP for some pdbs, and suffers from the same drawbacks as RASP on some proteins; structures for which RASP poorly predicts mutational effect are also, with a few exceptions, poorly predicted by our method. Yet, overall our performance is still significant, even more so when put in perspective with the fact that RASP regression is an ensemble model.

\begin{table}[h]
\centering

\begin{tabular}{p{1in}p{1.1in}p{1.1in}p{1.1in}p{1.1in}}
\\ \cline{2-5} 
\multicolumn{1}{c|}{\textbf{}}                & \multicolumn{1}{c|}{\textbf{Training Set}} & \multicolumn{1}{c|}{\textbf{Validation Set}} & \multicolumn{1}{c|}{\textbf{Test Set}} & \multicolumn{1}{c|}{\textbf{ThermoMutDB}} \\ \hline
\multicolumn{1}{|c|}{\textbf{Spearman r}}     & \multicolumn{1}{c|}{0.754}                 & \multicolumn{1}{c|}{0.518}                   & \multicolumn{1}{c|}{0.442}             & \multicolumn{1}{c|}{0.311}                \\ \hline
\multicolumn{1}{|c|}{\textbf{Pearson r}}      & \multicolumn{1}{c|}{0.758}                 & \multicolumn{1}{c|}{0.562}                   & \multicolumn{1}{c|}{0.412}             & \multicolumn{1}{c|}{0.251}                \\ \hline
\multicolumn{1}{|c|}{\textbf{RMSE}}           & \multicolumn{1}{c|}{0.794}                 & \multicolumn{1}{c|}{0.740}                   & \multicolumn{1}{c|}{0.935}             & \multicolumn{1}{c|}{2.288}                \\ \hline
\end{tabular}

\caption{Evaluation metrics for the $\Delta\Delta G$ scorer}
\centering
\label{Evaluation_scorer}
\end{table}

\begin{table}[h]
\centering
\begin{tabular}{p{1in}p{0.8in}p{0.8in}p{0.8in}p{0.8in}p{0.8in}p{0.8in}p{0.8in}p{0.8in}}
\\ \cline{1-9}
\multicolumn{1}{|c|}{\textbf{Number of mutations}}  & \multicolumn{1}{c|}{\textbf{1}} & \multicolumn{1}{c|}{\textbf{2}} & \multicolumn{1}{c|}{\textbf{3}} & \multicolumn{1}{c|}{\textbf{4}} & \multicolumn{1}{c|}{\textbf{5}} & \multicolumn{1}{c|}{\textbf{6}} & \multicolumn{1}{c|}{\textbf{7}} & \multicolumn{1}{c|}{\textbf{9}} \\ \hline
\multicolumn{1}{|c|}{\textbf{Spearman r}}  & \multicolumn{1}{c|}{0.349} & \multicolumn{1}{c|}{0.159} & \multicolumn{1}{c|}{0.381} & \multicolumn{1}{c|}{0.012} & \multicolumn{1}{c|}{0.271} & \multicolumn{1}{c|}{0.714} & \multicolumn{1}{c|}{-0.500} & \multicolumn{1}{c|}{1.000} \\ \hline
\multicolumn{1}{|c|}{\textbf{Pearson r}}  & \multicolumn{1}{c|}{0.342} & \multicolumn{1}{c|}{0.077} & \multicolumn{1}{c|}{0.346} & \multicolumn{1}{c|}{0.079} & \multicolumn{1}{c|}{0.202} & \multicolumn{1}{c|}{0.613} & \multicolumn{1}{c|}{-0.242} & \multicolumn{1}{c|}{1.000} \\ \hline
\multicolumn{1}{|c|}{\textbf{RMSE}}  & \multicolumn{1}{c|}{2.109} & \multicolumn{1}{c|}{2.571} & \multicolumn{1}{c|}{2.378} & \multicolumn{1}{c|}{1.972} & \multicolumn{1}{c|}{2.448} & \multicolumn{1}{c|}{1.965} & \multicolumn{1}{c|}{1.588} & \multicolumn{1}{c|}{0.810} \\ \hline
\end{tabular}
\caption{Metrics for our scorer across different numbers of mutations.}
\centering
\label{multi_mutation_metrics}
\end{table}

\begin{table}[h]
\centering
\begin{tabular}{p{1in}p{0.8in}p{0.8in}p{0.8in}p{0.8in}}
\\ \cline{1-5}
\multicolumn{1}{|c|}{\textbf{PDB ID}} & \multicolumn{1}{c|}{\textbf{1BNI}} & \multicolumn{1}{c|}{\textbf{1STN}} & \multicolumn{1}{c|}{\textbf{1VQB}} & \multicolumn{1}{c|}{\textbf{1RX4}} \\ \hline
\multicolumn{1}{|c|}{\textbf{Spearman r}} & \multicolumn{1}{c|}{0.503} & \multicolumn{1}{c|}{0.462} & \multicolumn{1}{c|}{0.519} & \multicolumn{1}{c|}{0.479} \\ \hline
\multicolumn{1}{|c|}{\textbf{Pearson r}} & \multicolumn{1}{c|}{0.456} & \multicolumn{1}{c|}{0.456} & \multicolumn{1}{c|}{0.523} & \multicolumn{1}{c|}{0.453} \\ \hline
\multicolumn{1}{|c|}{\textbf{RMSE}} & \multicolumn{1}{c|}{1.651} & \multicolumn{1}{c|}{1.632} & \multicolumn{1}{c|}{2.526} & \multicolumn{1}{c|}{1.193} \\ \hline
\end{tabular}
\caption{Scorer performance metrics on proteins with over 100 data points, as shown in Figure \ref{pdbid}.}
\centering
\label{pdb_mutation_metrics}
\end{table}

\section{Discussion}
These preliminary results show that the combination of decoupling of the atomic and residue scales, with the usage of an EGNN architecture, to allow flexibility on the number of mutations accessible to score, is promising.
In realising this exploratory work we faced two main challenges:
\begin{enumerate}[topsep=0pt,itemsep=1ex,partopsep=1ex,parsep=1ex]
\item The scorer had a tendency to over-fit the Mega-scale data-set.  
\item The current choice of a threshold for the residue graph is constrained. We ended up choosing 9{\AA}, where depending on the residues, a typical length for such interactions could go to 16{\AA} or more (for two tryptophans, given the max distance between their own atoms). But such a threshold would lead to a hyper-connected graph that would hinder the training. Generally speaking the graph building hyper-parameters, for example, the number of hops around the recovered nodes of interest (here one: neighbours one hop from the mutant nodes), would influence hyper-connectivity and our ability to not over-fit.
\end{enumerate}
We believe both of those problems, particularly the latter point, could be partially alleviated by finding a better encoding of meaningful distances of interaction, as well as including a more appropriate way to sum messages within the message passing loop \citep{ying2021centrality}.\\
In terms of evaluating scoring performance, when exploring the ThermoMutDB data-set as a potential out-of-distribution test set, we realised that Mega-scale has a significant advantage compared to all of the other available data-sets; it is experimentally consistent, both for \(\Delta\Delta G\) measurement and the use of AlphaFold for structure prediction. This is not the case for ThermoMutDB which is an aggregate of results obtained with a variety of methods for both stability measurement and structure determination, making it a challenge to understand why and how the model is failing to give accurate predictions. Training on such a data-set which will, for example, not include certain types of interactions, such as inter-domain interactions, as well as not containing the inherent real noise of protein structure prediction, is advantageous for its consistency and an inconvenience for its representational inaccuracies when compared to more "realistic" data-sets. 
In terms of computational performance, as we are using GNNs, we recognize that we lose an important aspect of the RASP model which is Rapid by naming. Yet, as the most time-consuming part is the construction of the residue sub-graph (roughly 5 seconds for subgraphs of less than 96 nodes with 8 CPUs, an A100 GPU and vectorization/jit features within JAX) saving it once and slightly modifying it to include the specific mutations later on, makes the model very efficient at assessing scores for multiple combinations of mutants within a pre-defined set of positions.\\
Finally, since we decoupled the atomic and the residue scales, it is now possible to swap elements from other successful models: for example ThermoNet. This exposes a new bottleneck, or rather a new further challenge, as it implies the creation of a new data-set including structures for each mutant present in the Mega-scale data-set. That would also have been the case if one wanted to include anti-symmetry properties within the model.

\section{Conclusion}
 In this work, we explored the possibility of using graph neural network models for scoring multiple substitution effects on protein stability. Our approach, based on the decoupling of atomic and residue scales by successively training two different scale-specific E-GNN models on massive experimental data-sets, shows promising results. Indeed, the model demonstrates an ability to predict effects of a variable number of mutations, even beyond what it has been trained on. Yet some key parameters of this modelling still need to be better understood; for example, a biologically reasonable edge distance threshold and an overall more appropriate way to handle connectivity in the created residue sub-graph. 

\bibliography{iclr2023_conference}

\begin{thebibliography}{30}
\providecommand{\natexlab}[1]{#1}
\providecommand{\url}[1]{\texttt{#1}}
\expandafter\ifx\csname urlstyle\endcsname\relax
  \providecommand{\doi}[1]{doi: #1}\else
  \providecommand{\doi}{doi: \begingroup \urlstyle{rm}\Url}\fi

\bibitem[Blaabjerg et~al.(2022)Blaabjerg, Kassem, Good, Jonsson, Cagiada,
  Johansson, Boomsma, Stein, and Lindorff-Larsen]{blaabjerg2022rapid}
Lasse~M Blaabjerg, Maher~M Kassem, Lydia~L Good, Nicolas Jonsson, Matteo
  Cagiada, Kristoffer~E Johansson, Wouter Boomsma, Amelie Stein, and Kresten
  Lindorff-Larsen.
\newblock Rapid protein stability prediction using deep learning
  representations.
\newblock \emph{bioRxiv}, pp.\  2022--07, 2022.

\bibitem[Bommarius et~al.(2006)Bommarius, Broering, Chaparro-Riggers, and
  Polizzi]{bommarius2006}
Andreas~S Bommarius, James~M Broering, Javier~F Chaparro-Riggers, and Karen~M
  Polizzi.
\newblock High-throughput screening for enhanced protein stability.
\newblock \emph{Current Opinion in Biotechnology}, 17\penalty0 (6):\penalty0
  606–610, Dec 2006.
\newblock ISSN 09581669.
\newblock \doi{10.1016/j.copbio.2006.10.001}.
\newblock URL
  \url{https://linkinghub.elsevier.com/retrieve/pii/S0958166906001455}.

\bibitem[Burley et~al.(2017)Burley, Berman, Kleywegt, Markley, Nakamura, and
  Velankar]{burley2017protein}
Stephen~K Burley, Helen~M Berman, Gerard~J Kleywegt, John~L Markley, Haruki
  Nakamura, and Sameer Velankar.
\newblock Protein data bank (pdb): the single global macromolecular structure
  archive.
\newblock \emph{Protein crystallography: methods and protocols}, pp.\
  627--641, 2017.

\bibitem[Cao et~al.(2019)Cao, Wang, He, Qi, and Zhang]{cao2019}
Huali Cao, Jingxue Wang, Liping He, Yifei Qi, and John~Z. Zhang.
\newblock Deepddg: Predicting the stability change of protein point mutations
  using neural networks.
\newblock \emph{Journal of Chemical Information and Modeling}, 59\penalty0
  (4):\penalty0 1508–1514, Apr 2019.
\newblock ISSN 1549-9596, 1549-960X.
\newblock \doi{10.1021/acs.jcim.8b00697}.
\newblock URL \url{https://pubs.acs.org/doi/10.1021/acs.jcim.8b00697}.

\bibitem[Case et~al.(2005)Case, Cheatham, Darden, Gohlke, Luo, Merz, Onufriev,
  Simmerling, Wang, and Woods]{case2005}
David~A. Case, Thomas~E. Cheatham, Tom Darden, Holger Gohlke, Ray Luo,
  Kenneth~M. Merz, Alexey Onufriev, Carlos Simmerling, Bing Wang, and Robert~J.
  Woods.
\newblock The amber biomolecular simulation programs.
\newblock \emph{Journal of Computational Chemistry}, 26\penalty0 (16):\penalty0
  1668–1688, Dec 2005.
\newblock ISSN 0192-8651, 1096-987X.
\newblock \doi{10.1002/jcc.20290}.
\newblock URL \url{https://onlinelibrary.wiley.com/doi/10.1002/jcc.20290}.

\bibitem[Das \& Baker(2008)Das and Baker]{das2008}
Rhiju Das and David Baker.
\newblock Macromolecular modeling with rosetta.
\newblock \emph{Annual Review of Biochemistry}, 77\penalty0 (1):\penalty0
  363–382, Jun 2008.
\newblock ISSN 0066-4154, 1545-4509.
\newblock \doi{10.1146/annurev.biochem.77.062906.171838}.
\newblock URL
  \url{https://www.annualreviews.org/doi/10.1146/annurev.biochem.77.062906.171838}.

\bibitem[Delaunay et~al.(2022)Delaunay, Fu, Bégué, McHardy, Djermani, Rooney,
  Tovchigrechko, Copoiu, Skwark, Lopez~Carraza, Lang, Beguir, and
  Şahin]{delaunay-etal-2022-gnn}
Antoine~P. Delaunay, Yunguan Fu, Alberto Bégué, Robert McHardy, Bachir~A.
  Djermani, Michael Rooney, Andrey Tovchigrechko, Liviu Copoiu, Marcin~J.
  Skwark, Nicolas Lopez~Carraza, Maren Lang, Karim Beguir, and Uğur Şahin.
\newblock Peptide-mhc structure prediction with mixed residue and atom graph
  neural network.
\newblock In \emph{Machine Learning in Structural Biology Workshop at the 36th
  Conference on Neural Information Processing Systems (NeurIPS)}, New Orleans,
  Louisiana, December 2022.
\newblock \doi{10.1101/2022.11.23.517618}.
\newblock URL \url{https://www.mlsb.io/}.

\bibitem[Folkman et~al.(2016)Folkman, Stantic, Sattar, and Zhou]{folkman2016}
Lukas Folkman, Bela Stantic, Abdul Sattar, and Yaoqi Zhou.
\newblock Ease-mm: Sequence-based prediction of mutation-induced stability
  changes with feature-based multiple models.
\newblock \emph{Journal of Molecular Biology}, 428\penalty0 (6):\penalty0
  1394–1405, Mar 2016.
\newblock ISSN 00222836.
\newblock \doi{10.1016/j.jmb.2016.01.012}.
\newblock URL
  \url{https://linkinghub.elsevier.com/retrieve/pii/S0022283616000310}.

\bibitem[Garcia~Satorras et~al.(2021)Garcia~Satorras, Hoogeboom, and
  Welling]{satoras2021}
Victor Garcia~Satorras, Emiel Hoogeboom, and Max Welling.
\newblock E(n) equivariant graph neural networks.
\newblock \emph{bioRxiv}, 2021.

\bibitem[Godoy-Ruiz et~al.(2004)Godoy-Ruiz, Perez-Jimenez, Ibarra-Molero, and
  Sanchez-Ruiz]{godoy-ruiz2004}
Raquel Godoy-Ruiz, Raul Perez-Jimenez, Beatriz Ibarra-Molero, and Jose~M.
  Sanchez-Ruiz.
\newblock Relation between protein stability, evolution and structure, as
  probed by carboxylic acid mutations.
\newblock \emph{Journal of Molecular Biology}, 336\penalty0 (2):\penalty0
  313–318, 2004.
\newblock \doi{10.1016/j.jmb.2003.12.048}.

\bibitem[Iqbal et~al.(2021)Iqbal, Li, Akutsu, Ascher, Webb, and
  Song]{iqbal2021}
Shahid Iqbal, Fuyi Li, Tatsuya Akutsu, David~B Ascher, Geoffrey~I Webb, and
  Jiangning Song.
\newblock Assessing the performance of computational predictors for estimating
  protein stability changes upon missense mutations.
\newblock \emph{Briefings in Bioinformatics}, 22\penalty0 (6):\penalty0
  bbab184, Nov 2021.
\newblock ISSN 1467-5463, 1477-4054.
\newblock \doi{10.1093/bib/bbab184}.
\newblock URL
  \url{https://academic.oup.com/bib/article/doi/10.1093/bib/bbab184/6289890}.

\bibitem[Jumper et~al.(2021)Jumper, Evans, Pritzel, Green, Figurnov,
  Ronneberger, Tunyasuvunakool, Bates, Žídek, Potapenko, Bridgland, Meyer,
  Kohl, Ballard, Cowie, Romera-Paredes, Nikolov, Jain, Adler, Back, Petersen,
  Reiman, Clancy, Zielinski, Steinegger, Pacholska, Berghammer, Bodenstein,
  Silver, Vinyals, Senior, Kavukcuoglu, Kohli, and Hassabis]{jumper2021}
John Jumper, Richard Evans, Alexander Pritzel, Tim Green, Michael Figurnov,
  Olaf Ronneberger, Kathryn Tunyasuvunakool, Russ Bates, Augustin Žídek, Anna
  Potapenko, Alex Bridgland, Clemens Meyer, Simon A.~A. Kohl, Andrew~J.
  Ballard, Andrew Cowie, Bernardino Romera-Paredes, Stanislav Nikolov, Rishub
  Jain, Jonas Adler, Trevor Back, Stig Petersen, David Reiman, Ellen Clancy,
  Michal Zielinski, Martin Steinegger, Michalina Pacholska, Tamas Berghammer,
  Sebastian Bodenstein, David Silver, Oriol Vinyals, Andrew~W. Senior, Koray
  Kavukcuoglu, Pushmeet Kohli, and Demis Hassabis.
\newblock Highly accurate protein structure prediction with alphafold.
\newblock \emph{Nature}, 596\penalty0 (7873):\penalty0 583–589, Aug 2021.
\newblock ISSN 0028-0836, 1476-4687.
\newblock \doi{10.1038/s41586-021-03819-2}.
\newblock URL \url{https://www.nature.com/articles/s41586-021-03819-2}.

\bibitem[Kawashima et~al.(2007)Kawashima, Pokarowski, Pokarowska, Kolinski,
  Katayama, and Kanehisa]{kawashima2007}
S.~Kawashima, P.~Pokarowski, M.~Pokarowska, A.~Kolinski, T.~Katayama, and
  M.~Kanehisa.
\newblock Aaindex: amino acid index database, progress report 2008.
\newblock \emph{Nucleic Acids Research}, 36:\penalty0 D202–D205, Dec 2007.
\newblock ISSN 0305-1048, 1362-4962.
\newblock \doi{10.1093/nar/gkm998}.
\newblock URL
  \url{https://academic.oup.com/nar/article-lookup/doi/10.1093/nar/gkm998}.

\bibitem[Kellogg et~al.(2011)Kellogg, Leaver-Fay, and Baker]{kellogg2011}
Elizabeth~H. Kellogg, Andrew Leaver-Fay, and David Baker.
\newblock Role of conformational sampling in computing mutation-induced changes
  in protein structure and stability: Conformational sampling in computing
  mutation-induced changes.
\newblock \emph{Proteins: Structure, Function, and Bioinformatics}, 79\penalty0
  (3):\penalty0 830–838, Mar 2011.
\newblock ISSN 08873585.
\newblock \doi{10.1002/prot.22921}.
\newblock URL \url{https://onlinelibrary.wiley.com/doi/10.1002/prot.22921}.

\bibitem[Li et~al.(2020)Li, Yang, Capra, and Gerstein]{li2020predicting}
Bian Li, Yucheng~T Yang, John~A Capra, and Mark~B Gerstein.
\newblock Predicting changes in protein thermodynamic stability upon point
  mutation with deep 3d convolutional neural networks.
\newblock \emph{PLoS computational biology}, 16\penalty0 (11):\penalty0
  e1008291, 2020.

\bibitem[Li et~al.(2021)Li, Panday, and Alexov]{li2021}
Gen Li, Shailesh~Kumar Panday, and Emil Alexov.
\newblock Saafec-seq: A sequence-based method for predicting the effect of
  single point mutations on protein thermodynamic stability.
\newblock \emph{International Journal of Molecular Sciences}, 22\penalty0
  (2):\penalty0 606, Jan 2021.
\newblock ISSN 1422-0067.
\newblock \doi{10.3390/ijms22020606}.
\newblock URL \url{https://www.mdpi.com/1422-0067/22/2/606}.

\bibitem[Marabotti et~al.(2021)Marabotti, Scafuri, and
  Facchiano]{marabotti2021}
Anna Marabotti, Bernardina Scafuri, and Angelo Facchiano.
\newblock Predicting the stability of mutant proteins by computational
  approaches: an overview.
\newblock \emph{Briefings in Bioinformatics}, 22\penalty0 (3):\penalty0
  bbaa074, May 2021.
\newblock ISSN 1467-5463, 1477-4054.
\newblock \doi{10.1093/bib/bbaa074}.
\newblock URL
  \url{https://academic.oup.com/bib/article/doi/10.1093/bib/bbaa074/5850907}.

\bibitem[Matthews(1993)]{matthews1993}
Brian~W. Matthews.
\newblock Structural and genetic analysis of protein stability.
\newblock \emph{Annual Review of Biochemistry}, 62\penalty0 (1):\penalty0
  139–160, Jun 1993.
\newblock ISSN 0066-4154, 1545-4509.
\newblock \doi{10.1146/annurev.bi.62.070193.001035}.
\newblock URL
  \url{https://www.annualreviews.org/doi/10.1146/annurev.bi.62.070193.001035}.

\bibitem[Pancotti et~al.(2021)Pancotti, Benevenuta, Repetto, Birolo, Capriotti,
  Sanavia, and Fariselli]{pancotti2021deep}
Corrado Pancotti, Silvia Benevenuta, Valeria Repetto, Giovanni Birolo, Emidio
  Capriotti, Tiziana Sanavia, and Piero Fariselli.
\newblock A deep-learning sequence-based method to predict protein stability
  changes upon genetic variations.
\newblock \emph{Genes}, 12\penalty0 (6):\penalty0 911, 2021.

\bibitem[Peng \& Alexov(2016)Peng and Alexov]{peng2016}
Yunhui Peng and Emil Alexov.
\newblock Investigating the linkage between disease-causing amino acid variants
  and their effect on protein stability and binding.
\newblock \emph{Proteins: Structure, Function, and Bioinformatics}, 84\penalty0
  (2):\penalty0 232–239, 2016.
\newblock \doi{10.1002/prot.24968}.

\bibitem[Qing et~al.(2022)Qing, Hao, Smorodina, Jin, Zalevsky, and
  Zhang]{qing2022}
Rui Qing, Shilei Hao, Eva Smorodina, David Jin, Arthur Zalevsky, and Shuguang
  Zhang.
\newblock Protein design: From the aspect of water solubility and stability.
\newblock \emph{Chemical Reviews}, 122\penalty0 (18):\penalty0 14085–14179,
  2022.
\newblock \doi{10.1021/acs.chemrev.1c00757}.

\bibitem[Schurmann et~al.(2011)Schurmann, Anton, Ivanov, Richter, Kuhn, and
  Walther]{schurmann2011}
Kathrin Schurmann, Monika Anton, Igor Ivanov, Constanze Richter, Hartmut Kuhn,
  and Matthias Walther.
\newblock Molecular basis for the reduced catalytic activity of the naturally
  occurring t560m mutant of human 12/15-lipoxygenase that has been implicated
  in coronary artery disease.
\newblock \emph{Journal of Biological Chemistry}, 286\penalty0 (27):\penalty0
  23920–23927, Jul 2011.
\newblock ISSN 00219258.
\newblock \doi{10.1074/jbc.M110.211821}.
\newblock URL
  \url{https://linkinghub.elsevier.com/retrieve/pii/S0021925819487437}.

\bibitem[Schymkowitz et~al.(2005)Schymkowitz, Borg, Stricher, Nys, Rousseau,
  and Serrano]{schymkowitz2005}
J.~Schymkowitz, J.~Borg, F.~Stricher, R.~Nys, F.~Rousseau, and L.~Serrano.
\newblock The foldx web server: an online force field.
\newblock \emph{Nucleic Acids Research}, 33:\penalty0 W382–W388, Jul 2005.
\newblock ISSN 0305-1048, 1362-4962.
\newblock \doi{10.1093/nar/gki387}.
\newblock URL
  \url{https://academic.oup.com/nar/article-lookup/doi/10.1093/nar/gki387}.

\bibitem[Stefl et~al.(2013)Stefl, Nishi, Petukh, Panchenko, and
  Alexov]{stefl2013}
Shannon Stefl, Hafumi Nishi, Marharyta Petukh, Anna~R. Panchenko, and Emil
  Alexov.
\newblock Molecular mechanisms of disease-causing missense mutations.
\newblock \emph{Journal of Molecular Biology}, 425\penalty0 (21):\penalty0
  3919–3936, Nov 2013.
\newblock ISSN 00222836.
\newblock \doi{10.1016/j.jmb.2013.07.014}.
\newblock URL
  \url{https://linkinghub.elsevier.com/retrieve/pii/S0022283613004464}.

\bibitem[Tsuboyama et~al.(2022)Tsuboyama, Dauparas, Chen, Laine, Behbahani,
  Weinstein, Mangan, Ovchinnikov, and Rocklin]{Tsuboyama2022}
Kotaro Tsuboyama, Justas Dauparas, Jonathan Chen, Elodie Laine, Yasser~M
  Behbahani, Jonathan~J Weinstein, Niall~M Mangan, Sergey Ovchinnikov, and
  Gabriel~J Rocklin.
\newblock Mega-scale experimental analysis of protein folding stability in
  biology and protein design.
\newblock \emph{bioRxiv}, 2022.

\bibitem[Wang et~al.(2021)Wang, Tang, Shan, and Zuo]{wang2021pros}
Shuyu Wang, Hongzhou Tang, Peng Shan, and Lei Zuo.
\newblock Pros-gnn: Predicting effects of mutations on protein stability using
  graph neural networks.
\newblock \emph{bioRxiv}, pp.\  2021--10, 2021.

\bibitem[Xavier et~al.(2021)Xavier, Nguyen, Karmarkar, Portelli, Rezende,
  Velloso, Ascher, and Pires]{xavier2021thermomutdb}
Joicymara~S Xavier, Thanh-Binh Nguyen, Malancha Karmarkar, Stephanie Portelli,
  P{\^a}mela~M Rezende, Joao~PL Velloso, David~B Ascher, and Douglas~EV Pires.
\newblock Thermomutdb: a thermodynamic database for missense mutations.
\newblock \emph{Nucleic acids research}, 49\penalty0 (D1):\penalty0 D475--D479,
  2021.

\bibitem[Xu et~al.(2020)Xu, Verma, Sheridan, Liaw, Ma, Marshall, McIntosh,
  Sherer, Svetnik, and Johnston]{xu2020}
Yuting Xu, Deeptak Verma, Robert~P. Sheridan, Andy Liaw, Junshui Ma,
  Nicholas~M. Marshall, John McIntosh, Edward~C. Sherer, Vladimir Svetnik, and
  Jennifer~M. Johnston.
\newblock Deep dive into machine learning models for protein engineering.
\newblock \emph{Journal of Chemical Information and Modeling}, 60\penalty0
  (6):\penalty0 2773–2790, Jun 2020.
\newblock ISSN 1549-9596, 1549-960X.
\newblock \doi{10.1021/acs.jcim.0c00073}.
\newblock URL \url{https://pubs.acs.org/doi/10.1021/acs.jcim.0c00073}.

\bibitem[Yang et~al.(2018)Yang, Urolagin, Niroula, Ding, Shen, and
  Vihinen]{yang2018}
Yang Yang, Siddhaling Urolagin, Abhishek Niroula, Xuesong Ding, Bairong Shen,
  and Mauno Vihinen.
\newblock Pon-tstab: Protein variant stability predictor. importance of
  training data quality.
\newblock \emph{International Journal of Molecular Sciences}, 19\penalty0
  (4):\penalty0 1009, Mar 2018.
\newblock ISSN 1422-0067.
\newblock \doi{10.3390/ijms19041009}.
\newblock URL \url{http://www.mdpi.com/1422-0067/19/4/1009}.

\bibitem[Ying et~al.(2021)Ying, Cai, Luo, Zheng, Ke, He, Shen, and
  Liu]{ying2021centrality}
Chengxuan Ying, Tianle Cai, Shengjie Luo, Shuxin Zheng, Guolin Ke, Di~He,
  Yanming Shen, and Tie-Yan Liu.
\newblock Do transformers really perform bad for graph representation?
\newblock \emph{bioRxiv}, 2021.

\end{thebibliography}
\bibliographystyle{iclr2023_conference}

\appendix
\section{Appendix}
\label{sec:appendix}
\subsection{data-set for atomic embedding (AE) training}
\begin{figure}
\centering
\includegraphics[width=0.99\linewidth]{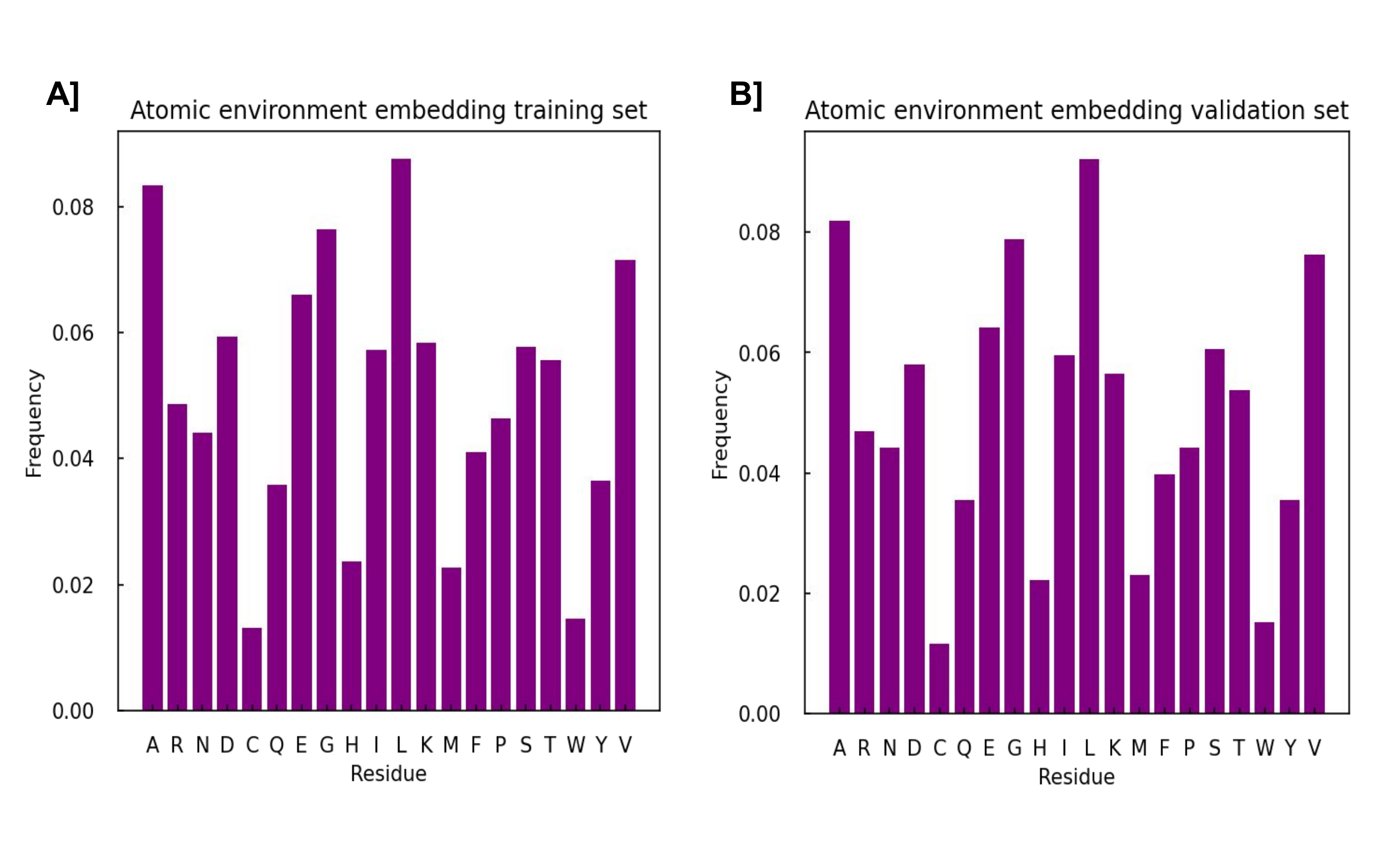}
\caption{Distribution of residue identities used to train our AE embedder.}
\label{freq_cav_data}
\centering
\end{figure}

As mentioned already we used the data-set provided by the RASP paper. We split the data by PDB id, keeping 90\% of the total pdbs for training and used the 10\% left for validation.\\
The pdbs kept for validation are the following:\\
'3VUP', '1A79', '3TFI', '1UXY', '1GNT', '3ZYZ', '2DEI', '4U9S', '4WGX', '4BGV', '1V70', '2HPA', '1KT5', '1BKP', '1FN9', '1LJ5', '4JZC', '1IJB', '2OI0', '4OO4', '5DNL', '1YB6', '5UL9', '1R4C', '4LD6', '5D6O', '1ZZG', '2O6S', '1MUN', '3KAN', '4AC1', '1GU7', '1LKT', '1G2B', '4PGG', '4CNP', '4FD9', '3NO0', '4HFQ', '1LDD', '1KWF', '1ZK9', '4BK7', '4R01', '2JA9', '4QXE', '1QEX', '2DP9', '4WFQ', '2WCU', '2I4A', '5BQ1', '5E4X', '3HCN', '1DAB', '1HBK', '2GAS', '1GNY', '1HTW', '1AOL', '5ML3', '2GGO', '1C5E', '3K1D', '1H2W', '2FC3', '2NWF', '1OIO', '4LWU', '5CMO', '4IY8', '3SBM', '3ZK1', '3D4M', '2BPS', '2D3Y', '4FD2', '5EG7', '2QJL', '1UZB', '4REL', '1POY', '1PGW', '1HJS', '4WHI', '1JKE', '5KB3', '2IXK', '1GK2', '2IJX', '3C8Y', '3POR', '1NAA', '1C39', '4WTO', '3G4H', '1A3H', '5EUV', '1UKU', '3SEB', '5H9I', '5N20', '5FXS', '3A4C', '1MIX', '5A7I', '1XNI', '4YTE', '2QCP', '1A9T', '4MIY', '1TYJ', '4I9A', '2Z5W', '8OHM', '5LQA', '2QXU', '5GMD', '1NHP', '1B87', '2OV0', '5EMI', '4PS2', '3TTY', '3UFC', '1H2E', '4F8X', '1UD2', '1T77', '1MV8', '1F2D', '1KEW', '4XQM', '5AN4', '1QUA', '2EBO', '3PJU', '3ZK9', '4GOC', '2Q9V', '5GZC', '2NQL', '5WS7', '2QIA', '1RKB', '4NVB', '3WR7', '4YHE', '2D58', '3E4G', '2CZQ', '1H70', '2R2D', '4A02', '3E9U', '1DD3', '2GDM', '1G6X', '3G5P', '2ECU', '2RKQ', '3TK0', '4OM8', '1UTG', '2JL1', '5FJQ', '1W53', '1HUL', '2QY1', '2CVI', '3NU1', '3RJP', '1UOY', '3Q46', '4MHP', '5LN4', '5KY2', '1LSL', '2H0M', '1ITX', '2YCI', '2FYG', '1F00', '1HJZ', '2A28', '3PGL', '2Z51', '5JDT', '1IT2', '3TDT', '1PCF', '5GJ6', '4O0A', '1EU4', '2QHT', '3LE2', '3OQT', '1LBV', '3TM0', '1KQP', '1LWD', '3IPJ', '2DSX', '1TJY', '2W5Q', '3KJH', '4FAZ', '1UCD', '1AO3', '2W72', '4LH9', '2EK0', '4EQS', '5HZ7', '2BEZ', '5MF5', '1CMB', '4ZPC', '2GWM', '2E9U', '1KTA', '1D2C', '1DCI', '4F6T', '1K0F', '5HZ8', '1LXY', '1QZM', '2F4M', '2JLQ', '2VBT', '1CFZ', '1N81'\\
Overall we ended up with: 945497 data points (over 2100 pdbs) for training and 107477 for validation (over 233 pdbs). The distributions of residues in both training and validation sets are drawn in Figure \ref{freq_cav_data}.

\begin{table}[]
\centering

\begin{tabular}{p{1in}p{1in}p{1in}p{1in}}
                                                                                             \\ \cline{2-4} 
\multicolumn{1}{c|}{\textbf{}}              & \multicolumn{1}{c|}{\textbf{RASP on}} & \multicolumn{1}{c|}{\textbf{This work on}} & \multicolumn{1}{c|}{\textbf{This work on}} \\ \multicolumn{1}{c|}{\textbf{}}              & \multicolumn{1}{c|}{\textbf{validation set}} & \multicolumn{1}{c|}{\textbf{validation set}} & \multicolumn{1}{c|}{\textbf{training set}} \\ \hline
\multicolumn{1}{|c|}{\textbf{F1 macro Average}}     & \multicolumn{1}{c|}{-}                               & \multicolumn{1}{c|}{0.61}                                 & \multicolumn{1}{c|}{0.63}              \\ \hline
\multicolumn{1}{|c|}{\textbf{Accuracy}}             & \multicolumn{1}{c|}{0.63}                            & \multicolumn{1}{c|}{0.65}                                 & \multicolumn{1}{c|}{0.67}              \\ \hline
\end{tabular}
\caption{Evaluation metrics for the AE embedder. Please note that the RASP validation set differs from our validation set as we used a different random seed.}
\centering
\label{Evaluation_Embedder}
\end{table}

\begin{table}[]
\centering
\begin{tabular}{c|c|c|c|}
\cline{2-4}
\multicolumn{1}{l|}{}                         & \textbf{\# pdbs} & \textbf{\# single mutation} & \textbf{\# double mutations} \\ \hline
\multicolumn{1}{|c|}{\textbf{Training set}}   & 215              & 284461                      & 136018                       \\ \hline
\multicolumn{1}{|c|}{\textbf{Validation set}} & 90               & 68050                       & 322                          \\ \hline
\multicolumn{1}{|c|}{\textbf{Test set}}       & 6                & 3451                        & 65                           \\ \hline
\end{tabular}
\caption{Statistic for the Mega-scale data-set subset used for training our scorer.}
\centering
\label{Table_mega_scale}
\end{table}

\subsection{Mega-scale data-set used}
\begin{figure}
\centering
\includegraphics[width=0.99\linewidth]{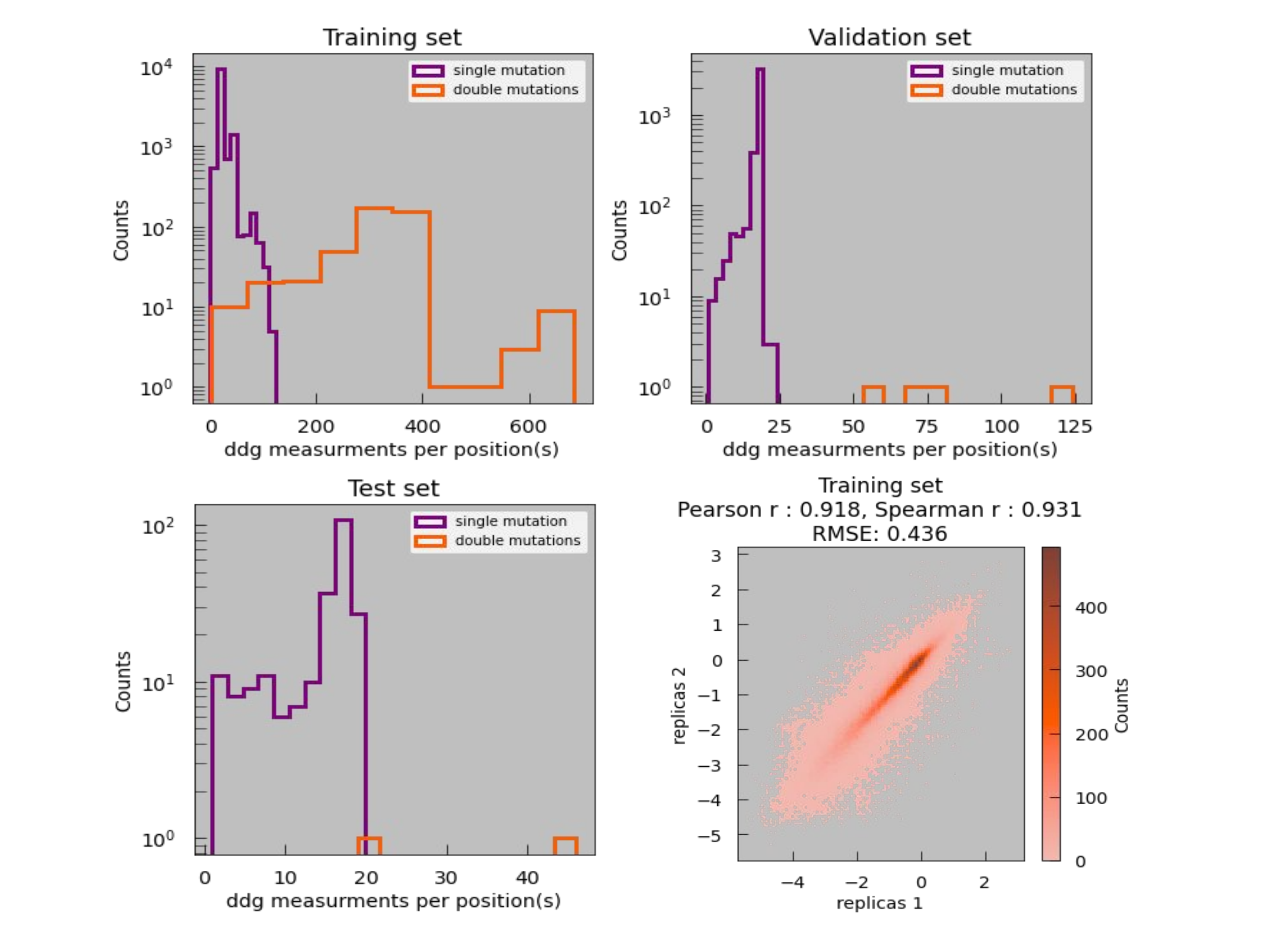}
\caption{Overview of the structure of our Mega-scale subset, in terms of mutational multiplicity at a given site as well as replica measurement consistency.}
\label{pos_degen}
\centering
\end{figure}

We split the Mega-scale data-set according to pdb ids (Table \ref{Table_mega_scale}). The Mega-scale data-set has some replica measurements in it (Figure \ref{pos_degen}), that we used as regularisation by including those different replica values during training. In general, in this data-set, multiple measurements will be performed at the same position but for different mutants (Figure \ref{pos_degen}).\\ \\
The pdbs kept within the distribution test (Mega-scale test set) are the following:\\'2M9I', '2N4T', '2YSC', '5AHT', '5XR0', '6YSE'.\\ \\
\flushleft{
The pdbs kept for validation are the following:

\begin{table}[h]
\flushleft

\begin{tabular}{p{1.268in}p{1.268in}p{1.268in}p{1.268in}}
\multicolumn{1}{c}{} & \multicolumn{1}{c}{} & \multicolumn{1}{c}{} & \multicolumn{1}{c}{} \\
'EEHEE-rd3-0602',& 'EEHEE-rd4-0371',& 'HEEH-KT-rd6-1415',& 'EEHEE-rd4-0003',\\
'EEHEE-rd4-0424',& 'HHH-rd4-0849',& 'HEEH-rd4-0097',& 'EEHEE-rd3-1702',\\
'HHH-rd1-0335',& 'EEHEE-rd4-0256',& 'EEHEE-rd3-1049',& 'EEHEE-rd4-0794',\\ 
'HEEH-rd4-0349',& 'HEEH-KT-rd6-0746',& 'EEHEE-rd4-0763',& 'EEHEE-rd3-1058',\\ 
'HHH-rd4-0870',& 'EEHEE-rd4-0647',& 'HHH-rd4-0124',& 'EEHEE-rd3-1498',\\
'HEEH-KT-rd6-0007',& 'HHH-rd1-0033',& 'HEEH-KT-rd6-0200',& 'HHH-rd2-0181',\\
'HHH-rd1-0473',& 'EEHEE-rd4-0481',& '2MWA',& 'HHH-rd1-0606',\\ 
'EEHEE-rd4-0469',& 'EEHEE-rd4-0363',& 'HEEH-KT-rd6-3632',& 'HEEH-KT-rd6-0790',\\ 
'HEEH-KT-rd6-0793',& 'HHH-rd1-0142',& 'HHH-rd1-0516',& 'EEHEE-rd4-0470',\\ 
'6FVC',& 'HHH-rd4-0557',& 'EEHEE-rd3-1627',& 'HHH-rd1-0578',\\ 
'HHH-rd2-0155',& 'HEEH-rd4-0094',& 'HHH-rd3-0008',& '2M8E',\\ 
'HHH-rd1-0196',& 'EEHEE-rd3-0094',& 'EEHEE-rd3-0146',& 'HHH-rd1-0598',\\ 
'2M8J',& 'EEHEE-rd3-1817',& 'EEHEE-rd3-1558',& 'HHH-rd4-0613',\\ 
'EEHEE-rd3-1367',& 'EEHEE-rd3-1587',& 'HEEH-rd3-0223',& 'EEHEE-rd4-0418',\\ 
'EEHEE-rd3-0657',& '5UYO',& 'EHEE-rd4-0300',& 'EEHEE-rd2-0770',\\ 
'EHEE-rd4-0195',& 'HHH-rd4-0816',& 'EEHEE-rd4-0784',& 'EEHEE-rd3-1810',\\ 
'EHEE-rd1-0101',& 'EHEE-rd2-0372',& 'EHEE-rd2-0196',& 'HHH-rd1-0756',\\ 
'EHEE-rd4-0463',& 'EHEE-rd1-0407',& 'EHEE-rd4-0325',& '2LX2',\\ 
'EHEE-rd4-0098',& 'EHEE-rd3-0067',& '2KCF',& 'EHEE-rd3-0053',\\ 
'EHEE-rd3-0035',& '2RRU',& 'EHEE-rd4-0340',& 'EHEE-rd4-0840',\\ 
'EHEE-rd4-0726',& 'EHEE-rd4-0172',& 'EHEE-rd4-0086',& 'EHEE-rd2-1257',\\ 
'EHEE-rd4-0394',& 'EHEE-rd2-0601',& 'EHEE-rd4-0044',& 'EHEE-rd2-0191',\\ 
'EHEE-rd2-0751',& 'EHEE-rd4-0864'.
\end{tabular}
\flushleft
\end{table}

    }

\subsection{ThermoMutDB data-set used} \label{app:thermomut}
\begin{figure}
\centering
\includegraphics[width=0.6\linewidth]{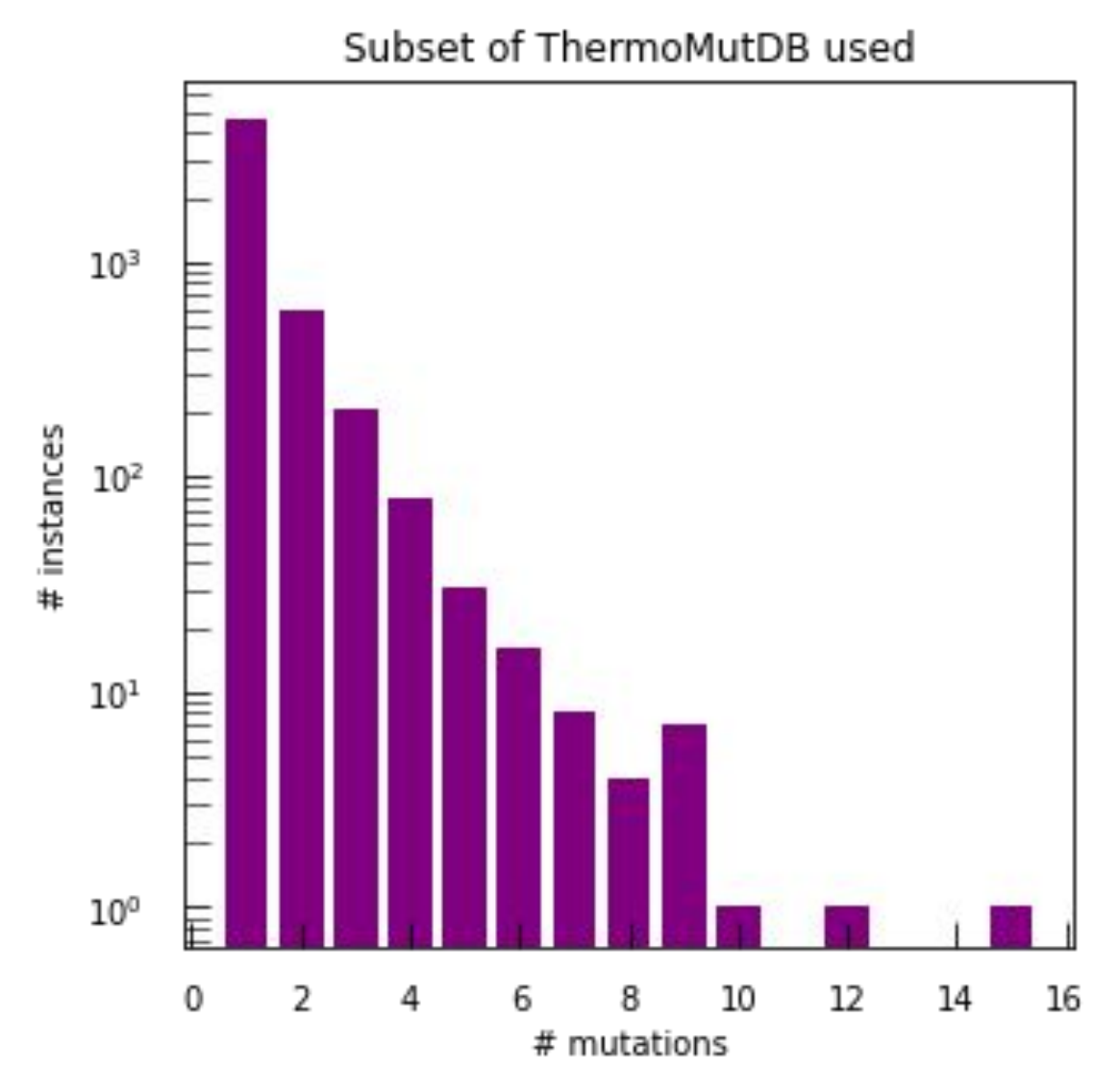}
\caption{Distribution of the number of mutations in the subset of ThermomMutDB that we used for testing.}
\label{Thermo_num_mut}
\centering
\end{figure}

We selected data points for which pdbs were retrievable, WT position in pdbs matched variant coding name, and $\Delta\Delta G$ was available. Whenever ThermoMutDB had replicates we merged them to their average value. Finally, we removed pdbs that were shared with the Mega-scale data-set. We thus ended up with 5514 data points over 322 pdbs and a range of multiple mutations as big as 15 (Figure \ref{Thermo_num_mut}).

\subsection{Figures and information related to the embedder model}
The architecture of the AE embedder is described in Table \ref{Table_embedder_model} and its training hyper-parameters in Table \ref{Table_hyper-param}.

\begin{table}[]
\centering
\begin{tabular}{p{1in}|p{1.1in}|p{1.1in}|p{1.1in}}
\cline{2-3}
                                                      & \textbf{Outputs size} & \textbf{Activation}                      &                                                                                                                           \\ \hline
\multicolumn{1}{|c|}{\textbf{$\Phi_e$}}               & {[}24,64{]}           & Swish                                    & \multicolumn{1}{c|}{\multirow{3}{*}{\begin{tabular}[c]{@{}c@{}}4 EGCL layers\\ Haiku implemented\\ MLP net\end{tabular}}} \\ \cline{1-3}
\multicolumn{1}{|c|}{\textbf{$\Phi_x$}}               & {[}32,1{]}            & Swish                                    & \multicolumn{1}{c|}{}                                                                                                     \\ \cline{1-3}
\multicolumn{1}{|c|}{\textbf{$\Phi_h$}}               & {[}24,64{]}           & Swish                                    & \multicolumn{1}{c|}{}                                                                                                     \\ \hline
\multicolumn{1}{|c|}{\textbf{Node features embedder}} & {[}64{]}      & None                              & \multicolumn{1}{c|}{\multirow{3}{*}{\begin{tabular}[c]{@{}c@{}}Haiku implemented\\ MLP net\end{tabular}}}                 \\ \cline{1-3}
\multicolumn{1}{|c|}{\textbf{Edge features embedder}} & {[}64{]}      & None                              & \multicolumn{1}{c|}{}                                                                                                     \\ \cline{1-3}
\multicolumn{1}{|c|}{\textbf{Pre scoring}}            & {[}64{]}  & None                              & \multicolumn{1}{c|}{}                                                                                                     \\ \hline
\multicolumn{1}{|c|}{\textbf{Output}}                 & {[}20{]}               & None (softmax included in loss function) & \multicolumn{1}{c|}{\begin{tabular}[c]{@{}c@{}}Haiku implemented\\ linear layer\end{tabular}}                             \\ \hline
\end{tabular}
\caption{Detailed view of the architecture of our EGNN AE. In a very Haiku way, only the output sizes of the different linear layers within the used MLP are presented. The naming of the modules from the EGCL layers follows \citep{satoras2021} naming. Other naming follows Figure\ref{EGNN}.}
\centering
\label{Table_embedder_model}
\end{table}

\begin{table}[]
\centering
\begin{tabular}{p{1in}|p{1.1in}|p{1.1in}|p{1.1in}}
\cline{2-3}
                                                      & \textbf{Outputs size} & \textbf{Activation}                      &                                                                                                                           \\ \hline
\multicolumn{1}{|c|}{\textbf{$\Phi_e$}}               & {[}6,8{]}           & Swish                                    & \multicolumn{1}{c|}{\multirow{3}{*}{\begin{tabular}[c]{@{}c@{}}2 EGCL layers\\ Haiku implemented\\ MLP net\end{tabular}}} \\ \cline{1-3}
\multicolumn{1}{|c|}{\textbf{$\Phi_x$}}               & {[}8,1{]}            & Swish                                    & \multicolumn{1}{c|}{}                                                                                                     \\ \cline{1-3}
\multicolumn{1}{|c|}{\textbf{$\Phi_h$}}               & {[}6,8{]}           & Swish                                    & \multicolumn{1}{c|}{}                                                                                                     \\ \hline
\multicolumn{1}{|c|}{\textbf{Node features embedder}} & {[}8{]}      & None                              & \multicolumn{1}{c|}{\multirow{3}{*}{\begin{tabular}[c]{@{}c@{}}Haiku implemented\\ MLP net\end{tabular}}}                 \\ \cline{1-3}
\multicolumn{1}{|c|}{\textbf{Edge features embedder}} & {[}8{]}      & None                              & \multicolumn{1}{c|}{}                                                                                                     \\ \cline{1-3}
\multicolumn{1}{|c|}{\textbf{Pre scoring}}            & {[}10{]}  & None                              & \multicolumn{1}{c|}{}                                                                                                     \\ \hline
\multicolumn{1}{|c|}{\textbf{Output}}                 & {[}1{]}              & sigmoid (Data is Fermi transformed) & \multicolumn{1}{c|}{\begin{tabular}[c]{@{}c@{}}Haiku implemented\\ linear layer\end{tabular}}                             \\ \hline
\end{tabular}
\caption{Detailed view of the architecture of our EGNN scorer. In a very Haiku way, only the output sizes of the different linear layers within the used MLP are presented. The naming of the modules from the EGCL layers follows \citep{satoras2021} naming. Other naming follows Figure\ref{EGNN}}
\centering
\label{Table_scorer_model}
\end{table}

\begin{figure}
\centering
\includegraphics[width=0.99\linewidth]{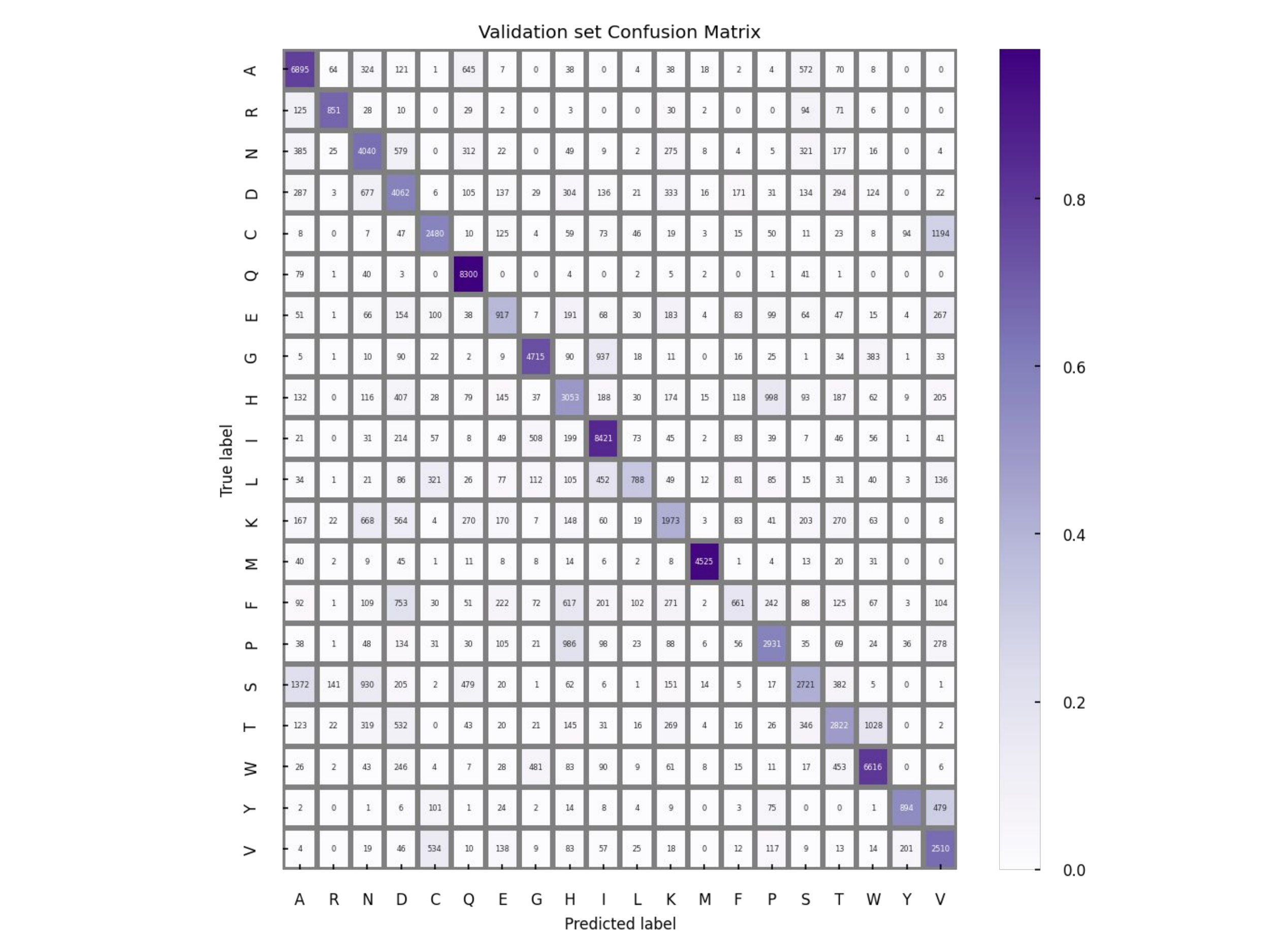}
\caption{Confusion matrix on the validation set for our embedder. Numbers inside the matrix correspond to counts and the color gradient corresponds to the fraction of predicted labels within a ground truth label category.}
\label{CM_val}
\centering
\end{figure}

\begin{table}[]
\centering
\begin{tabular}{p{1in}|p{1.1in}|p{1.1in}|}
\cline{2-3}
                                                            & \textbf{Embedder}       & \textbf{Scorer}        \\ \hline
\multicolumn{1}{|c|}{\textbf{Weight Decay (AdamW)}}         & $1e^{-6}$               & $1e^{-2}$              \\ \hline
\multicolumn{1}{|c|}{\textbf{Learning rate}}                & $3e^{-4}$               & $3e^{-5}$              \\ \hline
\multicolumn{1}{|c|}{\textbf{Batch Size}}                   & 96                      & 480                    \\ \hline
\multicolumn{1}{|c|}{\textbf{Max number of nodes}}                   & 500                     & 80                     \\ \hline
\multicolumn{1}{|c|}{\textbf{Max number of nodes in batch}} & $500*BatchSize$         & $80*BatchSize$         \\ \hline
\multicolumn{1}{|c|}{\textbf{Max number of edges in batch}} & $500^{1.5}*BatchSize$ & $80^{1.5}*BatchSize$ \\ \hline
\end{tabular}
\caption{Hyper-parameters used for both training our model and building the different graphs.}
\centering
\label{Table_hyper-param}
\end{table}

\subsection{Figures and information related to the scorer model}
The architecture of the scorer is described in Table \ref{Table_model} and its training hyper-parameters in Table \ref{Table_hyper-param}.

\begin{figure}
\centering
\includegraphics[width=0.99\linewidth]{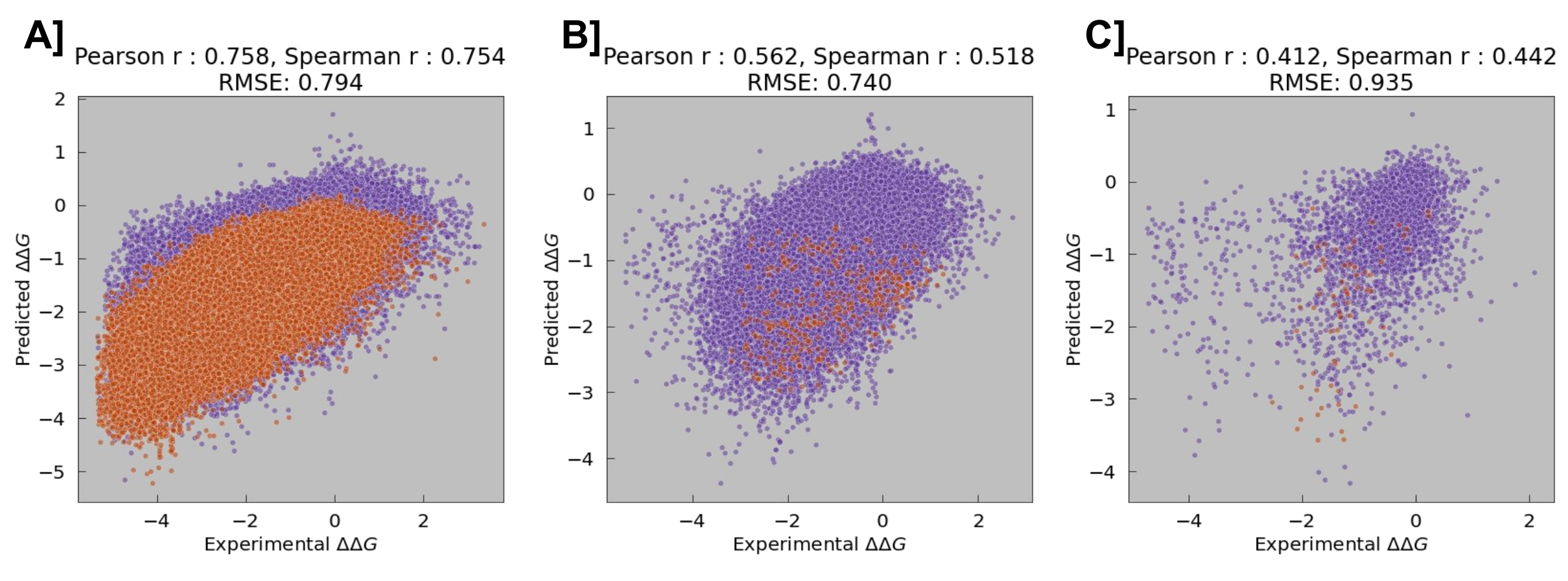}
\caption{Scorer model evaluation on training, validation and test set splits from Mega-scale. A] Mega-scale training set.  B] Mega-scale validation set.  C] Mega-scale test set. }
\label{Train_val_regress_pred}
\centering
\end{figure}

\begin{figure}
\centering
\includegraphics[width=0.99\linewidth]{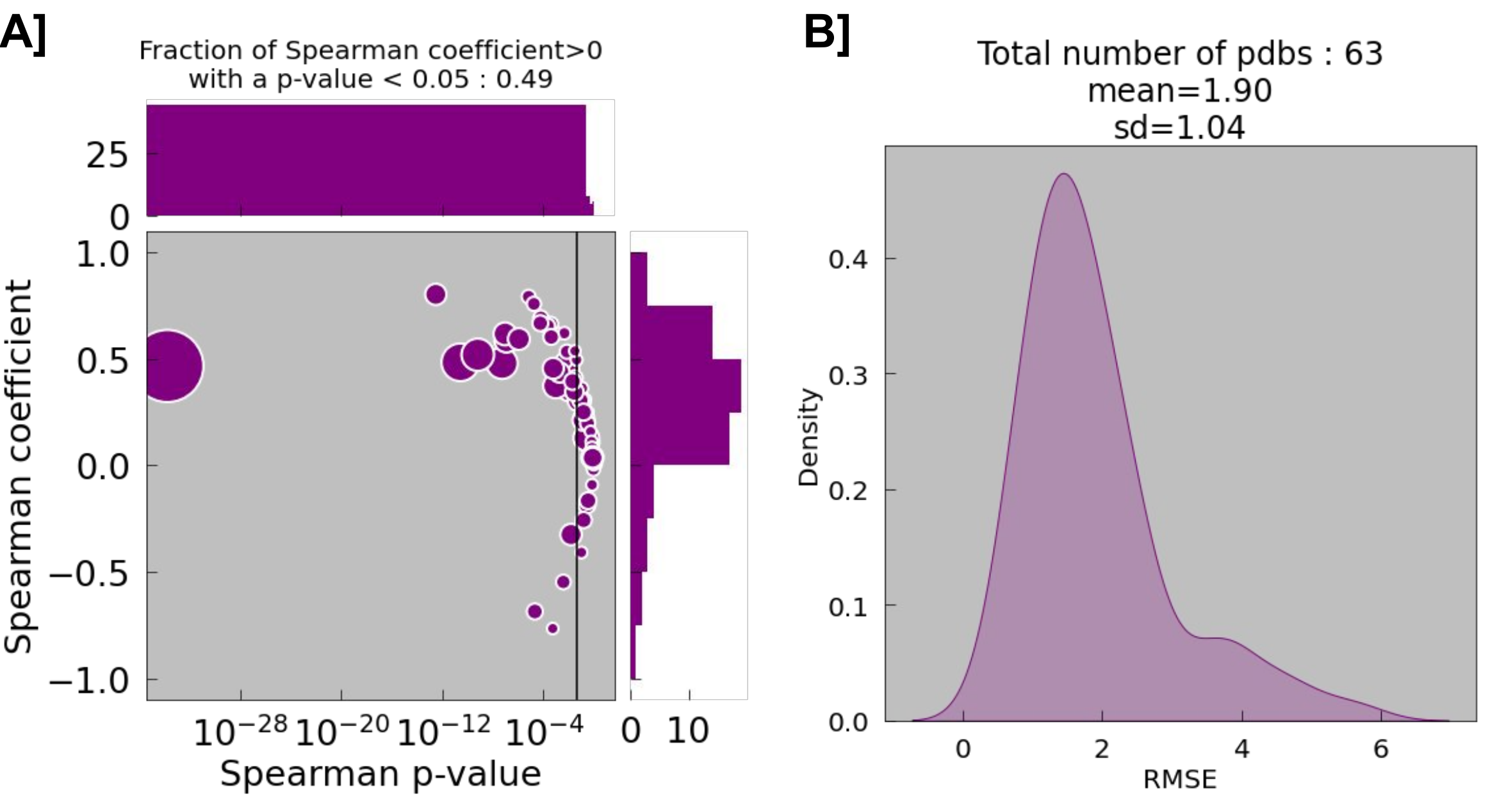}
\caption{Overview of the scorer performance over all the pdbs of ThermoMutDB having more than 15 data points. A] In this plot each point is a pair (Spearman r, Spearman p value) calculated per pdbs. Each point is thus a different pdb. Vertical black line indicates p value =0.05. Marker size is proportional to the number of points used in the correlation calculation B] Distribution of root mean squared error (RMSE) per pdb.}
\label{Distri_corr_pdbid}
\centering
\end{figure}
\subsection{ThermoMutDB subset for comparison performance with RASP}
The PDBs id used are : '1rtb', '1dkt', '4lyz', '1ycc', '1thq', '1rx4', '1ttg', '1stn', '1aps', '1sak', '3bdc', '2trx', '1cun', '1amq', '1n88', '1bta', '5emz', '3mbp', '2abd', '2brd', '1cey', '1hfy', '1bni', '1ris', '1arr', '2lzm', '2jie', '2afg', '1fkj', '1g3p', '5azu', '2pr5', '1h7m', '1lz1', '4hxj', '1rn1', '1ftg', '1bvc', '1bpi', '1bnz', '1igv', '1fc1', '1tup', '1l63', '1wq5', '1vqb', '1div', '1ten'
\begin{figure}
\centering
\includegraphics[width=0.99\linewidth]{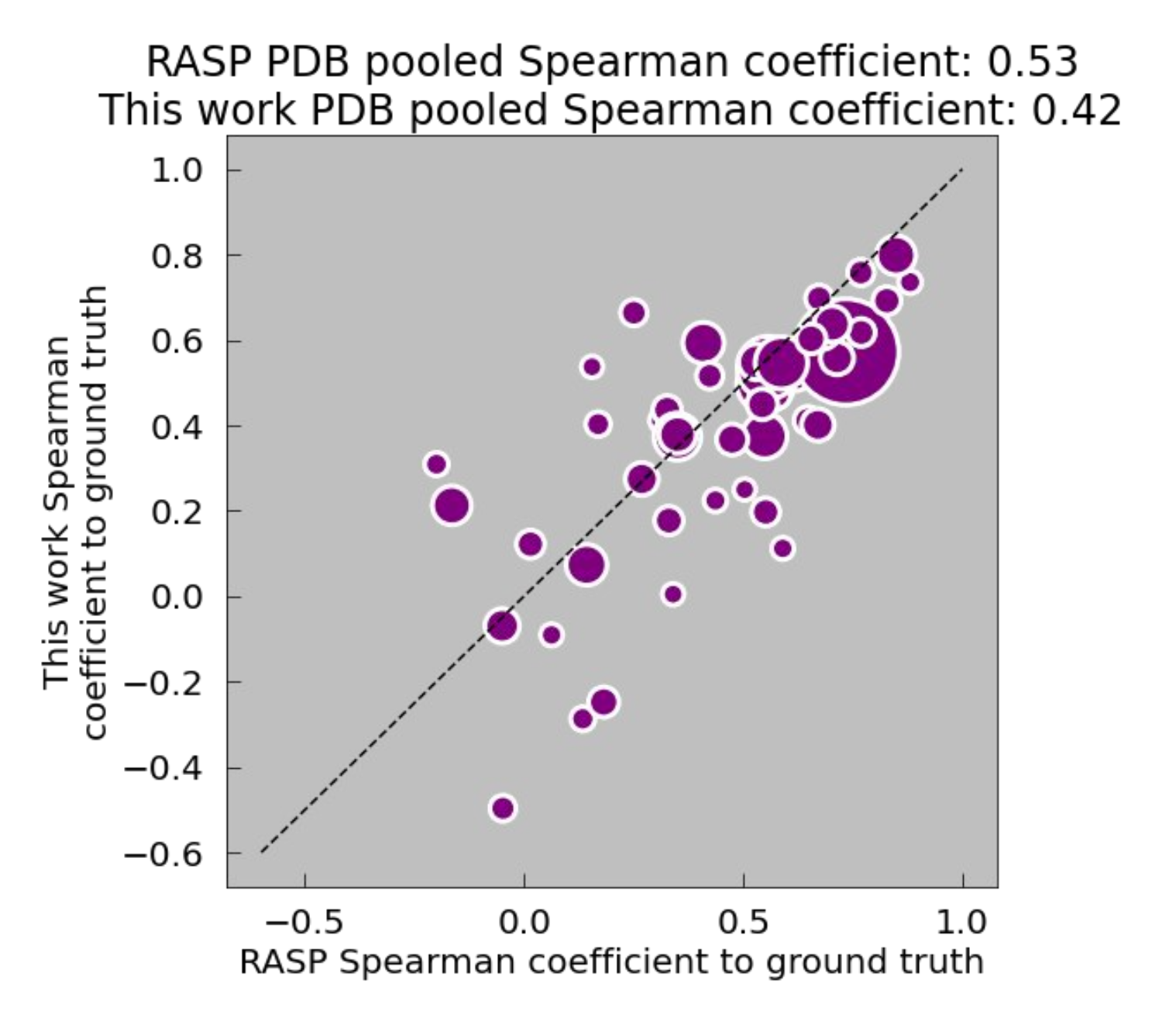}
\caption{Comparison between RASP and our work predicting on a subset of the ThermoMutDB data-set. Data points, and more specifically pdbs, were kept in the subset if a pdb had more than 15 data points with a single substitution mutation. Marker size is proportional to the number of points used for calculating the coefficient of correlation. Figure title describes the coefficient of correlation for the full subset : meaning without the breaking down by PDB ID.}
\label{RASP_comp}
\centering
\end{figure}
\end{document}